\documentclass[preprint,12pt]{elsarticle}

\usepackage{amssymb}
 \usepackage{amsthm}
\usepackage{latexsym}
\usepackage{amsmath}
\usepackage{amssymb}
\usepackage{diagbox}
\usepackage{slashbox}
\usepackage{color}
\usepackage{arydshln}

\newcommand{\udots}{\mathinner{\mskip1mu\raise1pt\vbox{\kern7pt\hbox{.}}
\mskip2mu\raise4pt\hbox{.}\mskip2mu\raise7pt\hbox{.}\mskip1mu}}
\usepackage{amsmath}

\newtheorem{theorem}{Theorem}

\newtheorem{lemma}{Lemma}

\begin{document}

\begin{frontmatter}



\title{ Multivariate tail covariance for generalized skew-elliptical distributions
}

\author{Baishuai  Zuo}
\author{Chuancun Yin\corref{cor1}}
\cortext[cor1]{Corresponding author.}
 \ead{ccyin@qfnu.edu.cn}

\address{School of Statistics, Qufu Normal University, Qufu, Shandong 273165, P. R. China}

\begin{abstract}
 In this paper, the multivariate tail covariance (MTCov) for generalized skew-elliptical distributions is considered. Some special cases for this  distribution, such as generalized skew-normal, generalized skew student-$t$, generalized skew-logistic and generalized skew-Laplace distributions, are also considered. In order to test the theoretical feasibility of our results, the MTCov for skewed and non
skewed normal distributions are computed and compared. Finally, we give a special formula of the MTCov for generalized skew-elliptical distributions.
\end{abstract}


\end{frontmatter}

\baselineskip=20pt

\section{Introduction}
 Furman and Landsman (2006) introduced tail variance (TV) measures of a random variable $X$, which is defined as
\begin{align*}
TV_{X}(x_{q})=E[(X-TCE_{X}(x_{q}))^{2}|X>x_{q}],
\end{align*}
where $TCE_{X}(x_{q})=E(X|X>x_{q})$ is tail conditional expectation of $X$ with a particular value $x_{q}$.
 Generally referred to as the $q$-th quantile with
$$x_{q}=\inf\{x\in\mathbb{R}|F_{X}(x)\geq q\}=\sup\{x\in\mathbb{R}|F_{X}(x)< q\},$$
for $q\in[0,~1]$, where $F_{X}(x)$ denotes distribution function of $X$ (see Denuit et al. (2005), p.17-18).
Recently, Landsman et al. (2018)  extended  TV measure to multivariate measure, and defined a novel type of a multivariate tail covariance (MTCov):
\begin{align}\label{(1)}
\nonumber&MTCov_{\boldsymbol{q}}(\mathbf{X})=E\left[(\mathbf{X}-MTCE_{\boldsymbol{q}}(\mathbf{X}))(\mathbf{X}-MTCE_{\boldsymbol{q}}(\mathbf{X}))^{T}|\mathbf{X}>VaR_{\boldsymbol{q}}(\mathbf{X})\right]\\
&=\inf_{\boldsymbol{c}\in\mathbb{R}^{n}}E\left[(\mathbf{X}-\boldsymbol{c})(\mathbf{X}-\boldsymbol{c})^{T}|\mathbf{X}>VaR_{\boldsymbol{q}}(\mathbf{X})\right].
\end{align}
Here
\begin{align}\label{(2)}
\nonumber&MTCE_{\boldsymbol{q}}(\mathbf{X})=E\left[\mathbf{X}|\mathbf{X}>VaR_{\boldsymbol{q}}(\mathbf{X})\right]\\
&=E[\mathbf{X}|X_{1}>VaR_{q_{1}}(X_{1}),\cdots,X_{n}>VaR_{q_{n}}(X_{n})],~\boldsymbol{q}=(q_{1},\cdots,q_{n})\in(0,~1)^{n},
\end{align} is multivariate tail conditional expectation (MTCE) of vector $\mathbf{X}$,
 where \\$\mathbf{X}=(X_{1},~X_{2},\cdots,X_{n})^{T}$ is an $n\times1$ vector of risks with cumulative distribution function (cdf) $F_{\mathbf{X}}(\boldsymbol{x})$ and tail function $\overline{F}_{\mathbf{X}}(\boldsymbol{x})$, $$VaR_{\boldsymbol{q}}(\mathbf{X})=(VaR_{q_{1}}(X_{1}),~VaR_{q_{2}}(X_{2}),\cdots,VaR_{q_{n}}(X_{n}))^{T},$$ and $VaR_{q_{k}}(X_{k}),~k=1,~2,\cdots,n$ is the value at risk (VaR) measure of the random variable $X_{k}$, being the $q_{k}$-th quantile of $X_{k}$. For the MTCE of some distributions, such as elliptical,  scale
mixtures of skew-normal, generalized skew-elliptical distributions,  we can see Landsman et al. (2016),  Mousavi et al. (2019) and Zuo and Yin (2020), respectively.

In Eini and Khaloozadeh (2020), authors derived expression of TV for generalized skew-elliptical distributions. Inspired by this, we derived MTCov for generalized skew-elliptical distributions, and presented expressions of MTCov for some special cases of this distributions, including generalized skew-normal, generalized skew student-$t$, generalized skew logistic, generalized skew Laplace distributions.

The rest of the paper is organized as follows. Section 2 introduces the definitions and properties of the generalized skew-elliptical distributions. In Section 3, we derive multivariate tail covariance for generalized skew-elliptical distributions. Some special cases are shown in Section 4. We present numerical illustration in Section 5. Finally, in Section 6,  is the concluding remarks.

\section{Generalized skew-elliptical distributions}
 A random vector $\mathbf{Y}$ is called an $n$-dimensional generalized skew-elliptical random vector, and denoted by $\mathbf{Y}\sim GSE_{n}\left(\boldsymbol{\mu},~\boldsymbol{\Sigma},~g_{n},~H(\cdot)\right)$. If it's probability density function (pdf) exists, the form will be (see McNeil et al. (2005) and Adcock et al. (2019))
\begin{align}\label{(3)}
f_{\boldsymbol{Y}}(\boldsymbol{y})=\frac{2c_{n}}{\sqrt{|\boldsymbol{\Sigma}|}}g_{n}\left\{\frac{1}{2}(\boldsymbol{y}-\boldsymbol{\mu})^{T}\mathbf{\Sigma}^{-1}(\boldsymbol{y}-\boldsymbol{\mu})\right\}H\left(\mathbf{\Sigma}^{-\frac{1}{2}}(\boldsymbol{y}-\boldsymbol{\mu})\right),~\boldsymbol{y}\in\mathbb{R}^{n},
\end{align}
where
\begin{align}\label{(4)}
f_{\boldsymbol{X}}(\boldsymbol{x}):=\frac{c_{n}}{\sqrt{|\boldsymbol{\Sigma}|}}g_{n}\left\{\frac{1}{2}(\boldsymbol{x}-\boldsymbol{\mu})^{T}\mathbf{\Sigma}^{-1}(\boldsymbol{x}-\boldsymbol{\mu})\right\},~\boldsymbol{x}\in\mathbb{R}^{n},
\end{align}
is the density of $n$-dimensional elliptical random vector $\mathbf{X}\sim E_{n}(\boldsymbol{\mu},\boldsymbol{\Sigma},g_{n})$.
Here $\boldsymbol{\mu}$ is an $n\times1$ location vector, $\mathbf{\Sigma}$ is an $n\times n$ scale matrix, and $g_{n}(u)$, $u\geq0$, is the density generator of $\mathbf{X}$. This density generator satisfies the condition
\begin{align*}
\int_{0}^{\infty}s^{n/2-1}g_{n}(s)\mathrm{d}s<\infty,
\end{align*}
 and the normalizing constant  $c_n$ is given by
\begin{align}\label{(5)}
c_{n}=\frac{\Gamma(n/2)}{(2\pi)^{n/2}}\left[\int_{0}^{\infty}s^{n/2-1}g_{n}(s)\mathrm{d}s\right]^{-1}.
\end{align}
$H(\boldsymbol{x}),~\boldsymbol{x}\in\mathbb{R}^{n},$ is called the skewing function satisfying $H(\boldsymbol{-x})=1-H(\boldsymbol{x})$ and $0\leq H(\boldsymbol{x})\leq1$.
The characteristic function of $\mathbf{X}$ takes the form  $\varphi_{\boldsymbol{X}}(\boldsymbol{t})=\exp\left\{i\boldsymbol{t}^{T}\boldsymbol{\mu}\right\}\psi\left(\frac{1}{2}\boldsymbol{t}^{T}\boldsymbol{\Sigma}\boldsymbol{t}\right),~\boldsymbol{t}\in \mathbb{R}^{n}$, with function $\psi(t):[0,\infty)\rightarrow\mathbb{R},$ called the characteristic generator (see Fang et al. (1990)).
Cumulative generator $\overline{G}_{n}(u)$ and $\overline{\mathcal{G}}_{n}(u)$ are defined as follows:
\begin{align*}
\overline{G}_{n}(u)=\int_{u}^{\infty}{g}_{n}(v)\mathrm{d}v
\end{align*}
and
\begin{align*}
\overline{\mathcal{G}}_{n}(u)=\int_{u}^{\infty}{G}_{n}(v)\mathrm{d}v,
\end{align*}
and their normalizing constants are, respectively, written (see Zuo et al. (2021)):
\begin{align}\label{(6)}
c_{n}^{\ast}=\frac{\Gamma(n/2)}{(2\pi)^{n/2}}\left[\int_{0}^{\infty}s^{n/2-1}\overline{G}_{n}(s)\mathrm{d}s\right]^{-1}
\end{align}
and
\begin{align}\label{(7)}
c_{n}^{\ast\ast}=\frac{\Gamma(n/2)}{(2\pi)^{n/2}}\left[\int_{0}^{\infty}s^{n/2-1}\overline{\mathcal{G}}_{n}(s)\mathrm{d}s\right]^{-1}.
\end{align}
 Shifted cumulative generator is also defined
\begin{align}\label{(8)}
{\overline{G}}_{n-1}^{\ast}(u)=\int_{u}^{\infty}g_{n}(v+a)\mathrm{d}v,~a\geq0,~n>1,
\end{align}
with ${\overline{G}}_{n-1}^{\ast}(u)<\infty$ (see Landsman et al. (2016)).
 \section{Multivariate tail covariance for generalized skew-elliptical distributions}
 Assume a random vector  $\mathbf{Y}\sim GSE_{n}\left(\boldsymbol{\mu},~\boldsymbol{\Sigma},~g_{n},~H(\cdot)\right)$ with finite vector $\boldsymbol{\mu}=(\mu_1,\cdots,\mu_n)^{T}$, positive defined matrix ${\bf {\Sigma}}= (\sigma_{ij})_{i,j=1}^{n}$ and probability density function $f_{\boldsymbol{Y}}(\boldsymbol{y})$.

$\mathbf{X}^{\ast}\sim E_{n}(\boldsymbol{\mu},~\boldsymbol{\Sigma},~\overline{G}_{n})$ and $\mathbf{X}^{\ast\ast}\sim E_{n}(\boldsymbol{\mu},~\boldsymbol{\Sigma},~\overline{\mathcal{G}}_{n})$ (see Zuo et al. (2021)) are respectively called elliptical random vectors with generators $\overline{G}_{n}(u)$ and $\overline{\mathcal{G}}_{n}(u)$, if their density functions (if them exist) defined by
\begin{align}\label{(9)}
f_{\boldsymbol{X}^{\ast}}(\boldsymbol{x})=\frac{c_{n}^{\ast}}{\sqrt{|\boldsymbol{\Sigma}|}}\overline{G}_{n}\left\{\frac{1}{2}(\boldsymbol{x}-\boldsymbol{\mu})^{T}\mathbf{\Sigma}^{-1}(\boldsymbol{x}-\boldsymbol{\mu})\right\},~\boldsymbol{x}\in\mathbb{R}^{n}
\end{align}
\begin{align}\label{(10)}
f_{\boldsymbol{X}^{\ast\ast}}(\boldsymbol{x})=\frac{c_{n}^{\ast\ast}}{\sqrt{|\boldsymbol{\Sigma}|}}\overline{\mathcal{G}}_{n}\left\{\frac{1}{2}(\boldsymbol{x}-\boldsymbol{\mu})^{T}\mathbf{\Sigma}^{-1}(\boldsymbol{x}-\boldsymbol{\mu})\right\},~\boldsymbol{x}\in\mathbb{R}^{n}.
\end{align}
So
 $\mathbf{Y}^{\ast}\sim GSE_{n}(\boldsymbol{\mu},~\mathbf{\Sigma},~\overline{G}_{n},~H(\cdot))$ and $\mathbf{Y}^{\ast\ast}\sim GSE_{n}(\boldsymbol{\mu},~\mathbf{\Sigma},~\overline{\mathcal{G}}_{n},~H(\cdot))$ are corresponding generalized skew-elliptical random vectors.\\
Let $\mathbf{Z}=\mathbf{\Sigma}^{-\frac{1}{2}}(\mathbf{Y}-\boldsymbol{\mu})\sim GSE_{n}\left(\boldsymbol{0},~\boldsymbol{I_{n}},~g_{n},~H(\cdot)\right).$
Writing
$$\boldsymbol{\lambda_{q}}=\left(\lambda_{1,\boldsymbol{q}},~\lambda_{2,\boldsymbol{q}},\cdots,\lambda_{n,\boldsymbol{q}}\right)^{T}=\mathbf{\Sigma}^{-\frac{1}{2}}(\boldsymbol{y_{q}-\mu}),$$
where $\boldsymbol{y_{q}}=VaR_{\boldsymbol{q}}(\boldsymbol{Y})$, $\boldsymbol{\lambda}_{-k,\boldsymbol{q}}=\left(\lambda_{1,\boldsymbol{q}},~\lambda_{2,\boldsymbol{q}},\cdots,\lambda_{k-1,\boldsymbol{q}},~\lambda_{k+1,\boldsymbol{q}},\cdots,\lambda_{n,\boldsymbol{q}}\right)^{T}$ and $\boldsymbol{\lambda}_{-kj,\boldsymbol{q}}=\left(\lambda_{1,\boldsymbol{q}},\cdots,\lambda_{k-1,\boldsymbol{q}},~\lambda_{k+1,\boldsymbol{q}},\cdots,\lambda_{j-1,\boldsymbol{q}},~\lambda_{j+1,\boldsymbol{q}},\cdots,\lambda_{n,\boldsymbol{q}}\right)^{T}$.

To derive formula for MTCov we define tail expectation $\overline{E}^{\boldsymbol{t}}_{\mathbf{Z}}[h(\boldsymbol{Z})]$ of $n$-dimensional random vector $\mathbf{Z}$ (see Zuo and Yin (2020) for details):
$$\overline{E}^{\boldsymbol{t}}_{\mathbf{Z}}[h(\boldsymbol{Z})]=\int_{\boldsymbol{t}}^{\infty}h(\boldsymbol{z})f_{\mathbf{Z}}(\boldsymbol{z})\mathrm{d}\boldsymbol{z},~~~\boldsymbol{z},\boldsymbol{t}\in\mathbb{R}^{n},$$
 where $h$ is an almost differentiable  function and $f_{\mathbf{Z}}(\boldsymbol{z})$ is pdf of $\mathbf{Z}$.\\
 So tail expectation $\overline{E}^{\boldsymbol{t}}_{\mathbf{W}_{-k}}[H(\boldsymbol{\xi}_{k,\boldsymbol{q}})]$ of $(n-1)$-dimensional elliptical random vector $$\mathbf{W}_{-k}=(W_{1},~W_{2},\cdots,W_{k-1},~W_{k+1},\cdots,W_{n})$$ enable to be expressed as
$$\overline{E}^{\boldsymbol{t}}_{\mathbf{W}_{-k}}[H(\boldsymbol{\xi}_{k,\boldsymbol{q}})]=\int_{\boldsymbol{t}}^{\infty}H(\boldsymbol{\xi}_{k,\boldsymbol{q}})f_{\mathbf{W}_{-k}}(\boldsymbol{w}_{-k})\mathrm{d}\boldsymbol{w}_{-k},~~~\boldsymbol{w}_{-k},\boldsymbol{t}\in\mathbb{R}^{n-1},$$ $~\mathrm{d}\boldsymbol{w}_{-k}=\mathrm{d}w_{1}\mathrm{d}w_{2}\cdots\mathrm{d}w_{k-1}\mathrm{d}w_{k+1}\cdots\mathrm{d}w_{n},$
with the pdf
\begin{align}\label{(11)}
\nonumber f_{\mathbf{W}_{-k}}(\boldsymbol{w}_{-k})&=c_{n-1,k}^{\ast}\overline{G}^{\ast}_{n-1,k}\left\{\frac{1}{2}\boldsymbol{w}_{-k}^{T}\boldsymbol{w}_{-k}\right\}\\
&=c_{n-1,k}^{\ast}\overline{G}_{n}\left\{\frac{1}{2}\boldsymbol{w}_{-k}^{T}\boldsymbol{w}_{-k}+\frac{1}{2}\lambda_{k,\boldsymbol{q}}^{2}\right\},~k=1,~2,\cdots,n,
\end{align}
where
 $\boldsymbol{\xi}_{k,\boldsymbol{q}}=(w_{1},~w_{2},\cdots,w_{k-1},~\lambda_{k,\boldsymbol{q}},~w_{k+1},\cdots,w_{n})^{T}$, and $\overline{G}^{\ast}_{n-1,k}$ is defined by (\ref{(8)}). In addition, $c_{n-1,k}^{\ast}$ is the normalizing constant of $\overline{G}^{\ast}_{n-1,k}$,
  $$c_{n-1,k}^{\ast}=\frac{\Gamma(\frac{n-1}{2})}{(2\pi)^{\frac{n-1}{2}}}\left[\int_{0}^{\infty}s^{\frac{n-1}{2}-1}\overline{G}_{n-1,k}^{\ast}(s)\mathrm{d}s\right]^{-1}.$$

We also define $f_{\mathbf{V}_{-k}}(\boldsymbol{v})$, $f_{\mathbf{U}_{-ij}}(\boldsymbol{u})$, pdf associated with elliptical random vectors $\mathbf{V}_{-k}$ and $\mathbf{U}_{-ij}$, respectively (see Landsman et al. (2018)):
\begin{align}\label{(12)}
f_{\mathbf{V}_{-k}}(\boldsymbol{v})=c_{n-1,k}^{\ast\ast}\overline{\mathcal{G}}_{n}\left\{\frac{1}{2}\boldsymbol{v}^{T}\boldsymbol{v}+\frac{1}{2}\lambda_{k,\boldsymbol{q}}^{2}\right\},~\boldsymbol{v}\in \mathbb{R}^{n-1},
\end{align}
\begin{align}\label{(13)}
f_{\mathbf{U}_{-ij}}(\boldsymbol{u})=c_{n-2,ij}^{\ast\ast}\overline{\mathcal{G}}_{n}\left\{\frac{1}{2}\boldsymbol{u}^{T}\boldsymbol{u}+\frac{1}{2}\lambda_{i,\boldsymbol{q}}^{2}+\frac{1}{2}\lambda_{j,\boldsymbol{q}}^{2}\right\},~\boldsymbol{u}\in \mathbb{R}^{n-2},
\end{align}
where $c_{n-1,k}^{\ast\ast}$
and
$c_{n-2,ij}^{\ast\ast}$ are corresponding normalizing constants of $f_{\mathbf{V}_{-k}}(\boldsymbol{v})$ and $f_{\mathbf{U}_{-ij}}(\boldsymbol{u})$.

We now rewrite Theorem 2 of Zuo and Yin (2020) as following lemma.
\begin{lemma}\label{le.1}
Assume that a random vector $\mathbf{Y}\sim GSE_{n}(\boldsymbol{\mu},~\mathbf{\Sigma},~g_{n},~H(\cdot))$ follows an $n$-variate $(n\geq2)$ generalized skew-elliptical distribution with probability density function $(\ref{(3)})$. Suppose
\begin{align}\label{(14)}
\lim_{\| \boldsymbol{z}\|\rightarrow +\infty}H(\boldsymbol{z})\overline{G}_{n}\left(\frac{1}{2}\boldsymbol{z}^{T}\boldsymbol{z}\right)=0,
\end{align}
where $\|\cdot\|$ denotes the Euclidean norm on $\mathbb{R}^{n}$.
 Then
\begin{align}\label{(15)} &MTCE_{\mathbf{Y}}(\boldsymbol{y_{q}})=\boldsymbol{\mu}+\mathbf{\Sigma}^{\frac{1}{2}}\frac{\boldsymbol{\delta_{\boldsymbol{q}}}}{\overline{F}_{\mathbf{Z}}(\boldsymbol{\lambda}_{\boldsymbol{q}})},
\end{align}
where
$$\boldsymbol{\delta_{\boldsymbol{q}}}=\left(\delta_{1,\boldsymbol{q}},~\delta_{2,\boldsymbol{q}},\cdots,\delta_{n,\boldsymbol{q}}\right)^{T}$$
with
\begin{align*}
\delta_{k,\boldsymbol{q}}=&\frac{2c_{n}}{c_{n-1,k}^{\ast}}\overline{E}^{\boldsymbol{\lambda}_{-k,\boldsymbol{q}}}_{\boldsymbol{W}_{-k}}[H(\boldsymbol{\xi}_{k,\boldsymbol{q}})]+\frac{2c_{n}}{c_{n}^{\ast}}\overline{E}^{\boldsymbol{\lambda}_{\boldsymbol{q}}}_{\mathbf{X}^{\ast}}[\partial_{k}H(\boldsymbol{X}^{\ast})],~k=1,~2,\cdots,n,
 \end{align*}
 $\mathbf{Z}\sim GSE_{n}\left(\boldsymbol{0},~\boldsymbol{I_{n}},~g_{n},~H(\cdot)\right)$, $\mathbf{X}^{\ast}\sim E_{n}(\boldsymbol{0},~\mathbf{I_{n}},~\overline{G}_{n})$, and pdf of $\mathbf{W}_{-k}$ is the same as in (\ref{(11)}). Furthermore, $\partial_{k}H(\boldsymbol{x}^{\ast})=\frac{\partial H(\boldsymbol{x}^{\ast})}{\partial x^{\ast}_{k}}$.
\end{lemma}
Denoting
$$\boldsymbol{\eta}_{k,\boldsymbol{q}}=(v_{1},~v_{2},\cdots,v_{k-1},~\lambda_{k,\boldsymbol{q}},~v_{k+1},\cdots,v_{n})^{T}$$
and
$$\boldsymbol{\tau}_{ij,\boldsymbol{q}}=(u_{1},\cdots,u_{i-1},\lambda_{i,\boldsymbol{q}},~u_{i+1},\cdots,u_{j-1},\lambda_{j,\boldsymbol{q}},u_{j+1},\cdots,u_{n})^{T}.$$
 \\In the following, we formulate the theorem that gives multivariate tail covariance (MTCov) for generalized skew-elliptical distributions.
\begin{theorem}\label{th.1} Assume that a random vector $\mathbf{Y}\sim GSE_{n}(\boldsymbol{\mu},~\mathbf{\Sigma},~g_{n},~H(\cdot))$ follows an $n$-variate $(n\geq2)$ generalized skew-elliptical distribution with probability density function $(\ref{(3)})$. We suppose
\begin{align}\label{(16)}
\lim_{\|\boldsymbol{z}\|\rightarrow +\infty}\|\boldsymbol{z}\|^{1/2}H(\boldsymbol{z})\overline{G}_{n}\left(\frac{1}{2}\boldsymbol{z}^{T}\boldsymbol{z}\right)=0
\end{align}
and
\begin{align}\label{(17)}
\lim_{\|\boldsymbol{z}\|\rightarrow +\infty}H(\boldsymbol{z})\overline{\mathcal{G}}_{n}\left(\frac{1}{2}\boldsymbol{z}^{T}\boldsymbol{z}\right)=0,
\end{align}
where $\|\cdot\|$ denotes the Euclidean norm on $\mathbb{R}^{n}$.
 Then
\begin{align}\label{(18)} &MTCov_{\mathbf{Y}}(\boldsymbol{y_{q}})=\mathbf{\Sigma}^{\frac{1}{2}}\mathbf{\Omega}_{\boldsymbol{q}}\mathbf{\Sigma}^{\frac{1}{2}},
\end{align}
where
\begin{align*}
\mathbf{\Omega}_{\boldsymbol{q},ij}&= \frac{2}{\overline{F}_{\mathbf{Z}}(\boldsymbol{\lambda_{q}})}\bigg\{\frac{c_{n}}{c_{n-2,ij}^{\ast\ast}}\overline{E}^{\boldsymbol{\lambda}_{-ij,\boldsymbol{q}}}_{\mathbf{U}_{-ij}}[H(\boldsymbol{\tau}_{ij,\boldsymbol{q}})]+\frac{c_{n}}{c_{n-1,i}^{\ast\ast}}\overline{E}^{\boldsymbol{\lambda}_{-i,\boldsymbol{q}}}_{\mathbf{V}_{-i}}[\partial_{i}H(\boldsymbol{\eta}_{i,\boldsymbol{q}})]\\
&~~~+\frac{c_{n}}{c_{n-1,j}^{\ast\ast}}\overline{E}^{\boldsymbol{\lambda}_{-j,\boldsymbol{q}}}_{\mathbf{V}_{-j}}[\partial_{j}H(\boldsymbol{\eta}_{j,\boldsymbol{q}})]+\frac{c_{n}}{c_{n}^{\ast\ast}}\overline{E}^{\boldsymbol{\lambda}_{\boldsymbol{q}}}_{\mathbf{X}^{\ast\ast}}[\partial_{ij}H(\boldsymbol{X}^{\ast\ast})]\bigg\}\\
&~~~-MTCE_{\mathbf{Z}}(\boldsymbol{\lambda_{q}})_{i}MTCE_{\mathbf{Z}}(\boldsymbol{\lambda_{q}})_{j}
,~i\neq j
\end{align*}
and
\begin{align*}
\mathbf{\Omega}_{\boldsymbol{q},ii}&= \frac{2}{\overline{F}_{\mathbf{Z}}(\boldsymbol{\lambda_{q}})}\bigg\{\frac{c_{n}}{c_{n-1,i}^{\ast}}\lambda_{i,\boldsymbol{q}}\overline{E}^{\boldsymbol{\lambda}_{-i,\boldsymbol{q}}}_{\boldsymbol{W}_{-i}}[H(\boldsymbol{\xi}_{i,\boldsymbol{q}})] +\frac{c_{n}}{c_{n-1,i}^{\ast\ast}}\overline{E}^{\boldsymbol{\lambda}_{-i,\boldsymbol{q}}}_{\mathbf{V}_{-i}}[\partial_{i}H(\boldsymbol{\eta}_{i,\boldsymbol{q}})] \\ &~~~+\frac{c_{n}}{c_{n}^{\ast\ast}}\overline{E}^{\boldsymbol{\lambda}_{\boldsymbol{q}}}_{\mathbf{X}^{\ast\ast}}[\partial_{ii}H(\boldsymbol{X}^{\ast\ast})]\bigg\}
+\frac{c_{n}}{c_{n}^{\ast}}\frac{\overline{F}_{\mathbf{Z^{\ast}}}(\boldsymbol{\lambda_{q}})}{\overline{F}_{\mathbf{Z}}(\boldsymbol{\lambda_{q}})}-[MTCE_{\mathbf{Z}}(\boldsymbol{\lambda_{q}})_{i}]^{2}
,
\end{align*}
with
\begin{align}\label{(19)}
MTCE_{\mathbf{Z}}(\boldsymbol{\lambda_{q}})_{k}=&\frac{2}{\overline{F}_{\mathbf{Z}}(\boldsymbol{\lambda}_{\boldsymbol{q}})}\left\{\frac{c_{n}}{c_{n-1,k}^{\ast}}\overline{E}^{\boldsymbol{\lambda}_{-k,\boldsymbol{q}}}_{\boldsymbol{W}_{-k}}[H(\boldsymbol{\xi}_{k,\boldsymbol{q}})]+\frac{c_{n}}{c_{n}^{\ast}}\overline{E}^{\boldsymbol{\lambda}_{\boldsymbol{q}}}_{\mathbf{X}^{\ast}}[\partial_{k}H(\boldsymbol{X}^{\ast})]\right\},
\end{align}
 $~i,~j,~k=1,~2,\cdots,n,$ $\mathbf{Z}\sim GSE_{n}\left(\boldsymbol{0},~\boldsymbol{I_{n}},~g_{n},~H(\cdot)\right)$, \\$\mathbf{Z}^{\ast}\sim GSE_{n}\left(\boldsymbol{0},~\boldsymbol{I_{n}},~\overline{G}_{n},~H(\cdot)\right)$, $\mathbf{X}^{\ast\ast}\sim E_{n}(\boldsymbol{0},~\mathbf{I_{n}},~\overline{\mathcal{G}}_{n})$, and pdfs of $\mathbf{W}_{-k}$,
  $\mathbf{V}_{-k}$ and $\mathbf{U}_{-ij}$ are same as in (\ref{(11)}), (\ref{(12)}) and (\ref{(13)}), respectively. In addition, $\partial_{ij}H(\boldsymbol{z})=\frac{\partial^{2} H(\boldsymbol{z})}{\partial z_{i}\partial z_{j}}$.
\end{theorem}
\noindent Proof. By definition of MTCov and MTCE in (\ref{(1)}) and (\ref{(2)}), we have
\begin{align*}
&MTCov_{\boldsymbol{q}}(\mathbf{Y})=E\left[(\mathbf{Y}-MTCE_{\boldsymbol{q}}(\mathbf{Y}))(\mathbf{Y}-MTCE_{\boldsymbol{q}}(\mathbf{Y}))^{T}|\mathbf{Y}>VaR_{\boldsymbol{q}}(\mathbf{Y})\right]\\
&=E\left[\mathbf{Y}\mathbf{Y}^{T}|\mathbf{Y}>VaR_{\boldsymbol{q}}(\mathbf{Y})\right]-MTCE_{\boldsymbol{q}}(\mathbf{Y})MTCE_{\boldsymbol{q}}^{T}(\mathbf{Y})\\
&=E\left[\mathbf{Y}\mathbf{Y}^{T}|\mathbf{Y}>VaR_{\boldsymbol{q}}(\mathbf{Y})\right]-E\left[\mathbf{Y}|\mathbf{Y}>VaR_{\boldsymbol{q}}(\mathbf{Y})\right]E\left[\mathbf{Y}|\mathbf{Y}>VaR_{\boldsymbol{q}}(\mathbf{Y})\right]^{T}.
\end{align*}
Applying the transformation $\mathbf{Z}=\mathbf{\Sigma}^{-\frac{1}{2}}(\mathbf{Y}-\boldsymbol{\mu})$, then using basic algebraic calculations we obtain
\begin{align*}
MTCov_{\boldsymbol{q}}(\mathbf{Y})&=\mathbf{\Sigma}^{\frac{1}{2}}\{E[\boldsymbol{Z}\boldsymbol{Z}^{T}|\mathbf{Z}>\boldsymbol{\lambda_{q}}]-E[\boldsymbol{Z}|\mathbf{Z}>\boldsymbol{\lambda_{q}}]E[\boldsymbol{Z}|\mathbf{Z}>\boldsymbol{\lambda_{q}}]^{T}\}\mathbf{\Sigma}^{\frac{1}{2}}\\
&=\mathbf{\Sigma}^{\frac{1}{2}}\{E[\boldsymbol{Z}\boldsymbol{Z}^{T}|\mathbf{Z}>\boldsymbol{\lambda_{q}}]-MTCE_{\mathbf{Z}}(\boldsymbol{\lambda_{q}})MTCE^{T}_{\mathbf{Z}}(\boldsymbol{\lambda_{q}})\}\mathbf{\Sigma}^{\frac{1}{2}},
\end{align*}
where $\boldsymbol{\lambda_{q}}=\mathbf{\Sigma}^{-\frac{1}{2}}(VaR_{\boldsymbol{q}}(\mathbf{Y})-\boldsymbol{\mu})$.\\
 Note that
\begin{align*}
&E[Z_{i}Z_{j}|\mathbf{Z}>\boldsymbol{\lambda_{q}}]=\frac{1}{\overline{F}_{\mathbf{Z}}(\boldsymbol{\lambda_{q}})}\int_{\boldsymbol{\lambda_{q}}}^{+\infty}2z_{i}z_{j}H(\boldsymbol{z})g_{n}(\frac{1}{2}\boldsymbol{z}^{T}\boldsymbol{z})\mathrm{d}\boldsymbol{z}\\
&=\frac{1}{\overline{F}_{\mathbf{Z}}(\boldsymbol{\lambda_{q}})}\bigg\{\int_{\boldsymbol{\lambda}_{-i,\boldsymbol{q}}}^{+\infty}2z_{j}H(\boldsymbol{z}_{i,\boldsymbol{q}})\overline{G}_{n}\left(\frac{1}{2}\boldsymbol{z}_{-i}^{T}\boldsymbol{z}_{-i}+\lambda_{i,\boldsymbol{q}}^{2}\right)\mathrm{d}\boldsymbol{z}_{-i}\\
&~~~+\int_{\boldsymbol{\lambda}_{\boldsymbol{q}}}^{+\infty}2z_{j}\frac{\partial H(\boldsymbol{z})}{\partial z_{i}}\overline{G}_{n}(\frac{1}{2}\boldsymbol{z}^{T}\boldsymbol{z})\mathrm{d}\boldsymbol{z}\bigg\}\\
&=\frac{1}{\overline{F}_{\mathbf{Z}}(\boldsymbol{\lambda_{q}})}\bigg\{\int_{\boldsymbol{\lambda}_{-ij,\boldsymbol{q}}}^{+\infty}2H(\boldsymbol{z}_{i,j,\boldsymbol{q}})\overline{\mathcal{G}}_{n}\left(\frac{1}{2}\boldsymbol{z}_{-ij}^{T}\boldsymbol{z}_{-ij}+\lambda_{i,\boldsymbol{q}}^{2}+\lambda_{j,\boldsymbol{q}}^{2}\right)\mathrm{d}\boldsymbol{z}_{-ij}\\
&~~~+\int_{\boldsymbol{\lambda}_{-i,\boldsymbol{q}}}^{+\infty}2\frac{\partial H(\boldsymbol{z}_{i,\boldsymbol{q}})}{\partial z_{j}}\overline{\mathcal{G}}_{n}(\frac{1}{2}\boldsymbol{z}_{-i}^{T}\boldsymbol{z}_{-i}+\frac{1}{2}\lambda_{i,\boldsymbol{q}}^{2})\mathrm{d}\boldsymbol{z}_{-i}\\
&~~~+\int_{\boldsymbol{\lambda}_{-j,\boldsymbol{q}}}^{+\infty}2\frac{\partial H(\boldsymbol{z}_{j,\boldsymbol{q}})}{\partial z_{i}}\overline{\mathcal{G}}_{n}(\frac{1}{2}\boldsymbol{z}_{-j}^{T}\boldsymbol{z}_{-j}+\frac{1}{2}\lambda_{j,\boldsymbol{q}}^{2})\mathrm{d}\boldsymbol{z}_{-j}\\
&~~~+\int_{\boldsymbol{\lambda}_{\boldsymbol{q}}}^{+\infty}2\frac{\partial^{2} H(\boldsymbol{z})}{\partial z_{i}\partial z_{j}}\overline{\mathcal{G}}_{n}(\frac{1}{2}\boldsymbol{z}^{T}\boldsymbol{z})\mathrm{d}\boldsymbol{z}\bigg\},for ~i\neq j, ~n\geq2,
\end{align*}
where $\boldsymbol{z}_{ij,\boldsymbol{q}}^{T}=(z_{1},\cdots,z_{i-1},\lambda_{i,\boldsymbol{q}},z_{i+1},\cdots,z_{j-1},\lambda_{j,\boldsymbol{q}},z_{j+1},\cdots,z_{n})$\\
and $\boldsymbol{z}_{i,\boldsymbol{q}}^{T}=(z_{1},\cdots,z_{i-1},\lambda_{i,\boldsymbol{q}},z_{i+1},\cdots,z_{n})$, and the second and third equalities we have used  integration by parts, Eq. (\ref{(16)}) and Eq. (\ref{(17)}).

Similarly, Using integration by parts, Eq. (\ref{(16)}) and Eq. (\ref{(17)}) we get
\begin{align*}
E[Z_{i}^{2}|\mathbf{Z}>\boldsymbol{\lambda_{q}}]&=\frac{1}{\overline{F}_{\mathbf{Z}}(\boldsymbol{\lambda_{q}})}\int_{\boldsymbol{\lambda_{q}}}^{+\infty}2z_{i}^{2}H(\boldsymbol{z})g_{n}(\frac{1}{2}\boldsymbol{z}^{T}\boldsymbol{z})\mathrm{d}\boldsymbol{z}\\
&=\frac{1}{\overline{F}_{\mathbf{Z}}(\boldsymbol{\lambda_{q}})}\bigg\{\int_{\boldsymbol{\lambda}_{-i,\boldsymbol{q}}}^{+\infty}2\lambda_{i,\boldsymbol{q}}H(\boldsymbol{z}_{i,\boldsymbol{q}})\overline{G}_{n}\left(\frac{1}{2}\boldsymbol{z}_{-i}^{T}\boldsymbol{z}_{-i}+\lambda_{i,\boldsymbol{q}}^{2}\right)\mathrm{d}\boldsymbol{z}_{-i}\\
&~~~+\int_{\boldsymbol{\lambda}_{\boldsymbol{q}}}^{+\infty}2\left[H(\boldsymbol{z})+z_{i}\frac{\partial H(\boldsymbol{z})}{\partial z_{i}}\right]\overline{G}_{n}\left(\frac{1}{2}\boldsymbol{z}^{T}\boldsymbol{z}\right)\mathrm{d}\boldsymbol{z}\bigg\}\\
&=\frac{1}{\overline{F}_{\mathbf{Z}}(\boldsymbol{\lambda_{q}})}\bigg\{\int_{\boldsymbol{\lambda}_{-i,\boldsymbol{q}}}^{+\infty}2\lambda_{i,\boldsymbol{q}}H(\boldsymbol{z}_{i,\boldsymbol{q}})\overline{G}_{n}\left(\frac{1}{2}\boldsymbol{z}_{-i}^{T}\boldsymbol{z}_{-i}+\lambda_{i,\boldsymbol{q}}^{2}\right)\mathrm{d}\boldsymbol{z}_{-i}\\
&~~~+\int_{\boldsymbol{\lambda}_{\boldsymbol{q}}}^{+\infty}2H(\boldsymbol{z})\overline{G}_{n}\left(\frac{1}{2}\boldsymbol{z}^{T}\boldsymbol{z}\right)\mathrm{d}\boldsymbol{z}
\end{align*}
\begin{align*}
&~~~+\int_{\boldsymbol{\lambda}_{-i,\boldsymbol{q}}}^{+\infty}2\frac{\partial H(\boldsymbol{z}_{i,\boldsymbol{q}})}{\partial z_{i}}\overline{\mathcal{G}}_{n}(\frac{1}{2}\boldsymbol{z}_{-i}^{T}\boldsymbol{z}_{-i}+\frac{1}{2}\lambda_{i,\boldsymbol{q}}^{2})\mathrm{d}\boldsymbol{z}_{-i}\\
&~~~+\int_{\boldsymbol{\lambda}_{\boldsymbol{q}}}^{+\infty}2\frac{\partial^{2} H(\boldsymbol{z})}{\partial z_{i}^{2}}\overline{\mathcal{G}}_{n}(\frac{1}{2}\boldsymbol{z}^{T}\boldsymbol{z})\mathrm{d}\boldsymbol{z}\bigg\}.
\end{align*}
As for $MTCE_{\mathbf{Z}}(\boldsymbol{\lambda_{q}})_{k}$, using Lemma \ref{le.1} we immediately obtain (\ref{(19)}).
Therefore we obtain $(\ref{(18)})$, which completes the proof of Theorem $1$.\\
$\mathbf{Remark~1.}$ We observe that, when $H(\cdot)=\frac{1}{2}$ in Theorem $1$, one gets the formula of Theorem 2 in Landsman et al. (2018):
Its' form is the same as that in (\ref{(18)}), where
\begin{align*}
\mathbf{\Omega}_{\boldsymbol{q},ij}=&\frac{1}{\overline{F}_{\mathbf{Z}}(\boldsymbol{\lambda_{q}})}\bigg\{\frac{c_{n}}{c_{n-2,ij}^{\ast\ast}}\overline{F}_{\mathbf{U}_{-ij}}(\boldsymbol{\lambda}_{-ij,\boldsymbol{q}})\bigg\}-MTCE_{\mathbf{Z}}(\boldsymbol{\lambda_{q}})_{i}MTCE_{\mathbf{Z}}(\boldsymbol{\lambda_{q}})_{j},~i\neq j
\end{align*}
and
\begin{align*}
\mathbf{\Omega}_{\boldsymbol{q},ii}&=\frac{1}{\overline{F}_{\mathbf{Z}}(\boldsymbol{\lambda_{q}})}\bigg\{\lambda_{i,\boldsymbol{q}}\frac{c_{n}}{c_{n-1,i}^{\ast}}\overline{F}_{\boldsymbol{W}_{-i}}(\boldsymbol{\lambda}_{-i,\boldsymbol{q}}) \bigg\}
+\frac{c_{n}}{c_{n}^{\ast}}\frac{\overline{F}_{\mathbf{Z^{\ast}}}(\boldsymbol{\lambda_{q}})}{\overline{F}_{\mathbf{Z}}(\boldsymbol{\lambda_{q}})}-[MTCE_{\mathbf{Z}}(\boldsymbol{\lambda_{q}})_{i}]^{2}
,
\end{align*}
with
\begin{align*} &MTCE_{\mathbf{Z}}(\boldsymbol{\lambda_{q}})_{k}=\frac{c_{n}}{c_{n-1,k}^{\ast}}\frac{\overline{F}_{\boldsymbol{W}_{-k}}(\boldsymbol{\lambda}_{-k,\boldsymbol{q}})}{\overline{F}_{\mathbf{Z}}(\boldsymbol{\lambda}_{\boldsymbol{q}})},
\end{align*}
$~i,~j,~k=1,~2,\cdots,n,$ $\mathbf{Z}\sim E_{n}\left(\boldsymbol{0},~\boldsymbol{I_{n}},~g_{n}\right)$, $\mathbf{Z}^{\ast}\sim E_{n}\left(\boldsymbol{0},~\boldsymbol{I_{n}},~\overline{G}_{n}\right)$, and pdfs of $\mathbf{W}_{-k}$
   and $\mathbf{U}_{-ij}$ are same as those in Theorem 1.

\noindent $\mathbf{Remark~2.}$ When $n=1$, we obtain tail variance (TV) for generalized skew-elliptical distributions:
\begin{align*}
 &TV_{Y}(y_{q})=\sigma^{2}\Omega_{q},
\end{align*}
where
\begin{align*}
\Omega_{q}=&\frac{2}{\overline{F}_{Z}(\lambda_{q})}\bigg\{c_{1}\lambda_{q}H(\lambda_{q})\overline{G_{1}}(\frac{1}{2}\lambda_{q}^2)
+ c_{1}H'(\lambda_{q})\overline{\mathcal{G}_{1}}(\frac{1}{2}\lambda_{q}^{2})\\
&+\frac{c_{1}}{c_{1}^{\ast\ast}}\overline{E}^{\lambda_q}_{X^{\ast\ast}}[H''(X^{\ast\ast})]\bigg\}+\frac{c_{1}}{c_{1}^{\ast}}\frac{\overline{F}_{Z^{\ast}}(\lambda_{q})}{\overline{F}_{Z}(\lambda_{q})}-[TCE_{Z}(\lambda_{q})]^{2},
\end{align*}
with
\begin{align*}
 &TCE_{Z}(\lambda_{q})=2c_{1} H(\lambda_{q})\frac{\overline{G}_{1}\left(\frac{1}{2}\lambda_{q}^{2}\right)}{\overline{F}_{Z}(\lambda_{q})}+\frac{2c_{1}}{c_{1}^{\ast}}\frac{\overline{E}^{\lambda_{q}}_{X^{\ast}}[H'(X^{\ast})]}{\overline{F}_{Z}(\lambda_{q})}.
\end{align*}
It is coincide with the result of Theorem of  Eini and Khaloozadeh (2020).

From matrix MTCov we can obtain the Multivariate Tail Correlation
matrix (see Landsman et al. (2018)):
\begin{align}\label{(20)}
MTCorr_{\mathbf{Y}}=\bigg(\frac{MTCov_{\mathbf{Y},ij}}{\sqrt{MTCov_{\mathbf{Y},ii}}\sqrt{MTCov_{\mathbf{Y},jj}}}\bigg)_{ij=1,\cdots,n}.
\end{align}
\section{Special cases}
We now consider special cases of the generalized skew-elliptical distributions, such as generalized skew-normal, generalized skew student-$t$, generalized skew-logistic and generalized skew-Laplace distributions. Because their forms of MTCov are same as that in Theorem $1$, we only give $\mathbf{\Omega}_{\boldsymbol{q},ij},~i=j~and~ i\neq j$.\\
$\mathbf{Example~4.1}$ (Generalized skew-normal distribution). The density function of an
n-dimension generalized skew normal random vector $\mathbf{Y}$, with location parameter $\boldsymbol{\mu}$, scale
matrix $\mathbf{\Sigma}$ and skewing function $H(\cdot): \mathbb{R}\rightarrow\mathbb{R}$, is given by
\begin{align*}
f_{\boldsymbol{Y}}(\boldsymbol{y})=\frac{2}{\sqrt{|\mathbf{\Sigma}|}(2\pi)^{\frac{n}{2}}}\exp\left\{-\frac{1}{2}(\boldsymbol{y}-\boldsymbol{\mu})^{T}\mathbf{\Sigma}^{-1}(\boldsymbol{y}-\boldsymbol{\mu})\right\}
 H\left(\boldsymbol{\gamma}^{T}\mathbf{\Sigma}^{-\frac{1}{2}}(\boldsymbol{y}-\boldsymbol{\mu})\right),
\end{align*}
 $\boldsymbol{y}\in\mathbb{R}^{n},$ where $\boldsymbol{\gamma}=(\gamma_{1},~\gamma_{2},~\cdots,~\gamma_{n})^{T}\in\mathbb{R}^{n}$. We denote it by\\ $\mathbf{Y}\sim GSN_{n}(\boldsymbol{\mu},~\mathbf{\Sigma},~\boldsymbol{\gamma},~H(\cdot))$. In this case,\\ $\overline{\mathcal{G}}_{n}(u)=\overline{G}_{n}(u)=g_{n}(u)=\exp(-u)$, $c_{n}^{\ast\ast}=c_{n}^{\ast}=c_{n}=(2\pi)^{-\frac{n}{2}}$ and $$H\left(\mathbf{\Sigma}^{-\frac{1}{2}}(\boldsymbol{y}-\boldsymbol{\mu})\right)=H\left(\boldsymbol{\gamma}^{T}\mathbf{\Sigma}^{-\frac{1}{2}}(\boldsymbol{y}-\boldsymbol{\mu})\right).$$ Since
 $$f_{\mathbf{W}_{-k}}(\boldsymbol{w})=c_{n-1,k}^{\ast}\exp\left\{-\frac{1}{2}\boldsymbol{w}^{T}\boldsymbol{w}-\frac{1}{2}\lambda_{k,\boldsymbol{q}}^{2}\right\}=\phi_{n-1}(\boldsymbol{w}),~\boldsymbol{w}\in \mathbb{R}^{n-1},$$
 $$f_{\mathbf{V}_{-k}}(\boldsymbol{v})=c_{n-1,k}^{\ast\ast}\exp\left\{-\frac{1}{2}\boldsymbol{v}^{T}\boldsymbol{v}-\frac{1}{2}\lambda_{k,\boldsymbol{q}}^{2}\right\}=\phi_{n-1}(\boldsymbol{v}),~\boldsymbol{v}\in \mathbb{R}^{n-1},$$
$$f_{\mathbf{U}_{-ij}}(\boldsymbol{u})=c_{n-2,ij}^{\ast\ast}\exp\left\{\frac{1}{2}\boldsymbol{u}^{T}\boldsymbol{u}-\frac{1}{2}\lambda_{i,\boldsymbol{q}}^{2}-\frac{1}{2}\lambda_{j,\boldsymbol{q}}^{2}\right\}=\phi_{n-2}(\boldsymbol{u}),~\boldsymbol{u}\in \mathbb{R}^{n-2},$$
$\phi_{k}(\cdot)$ is the pdf of $k$-dimensional standard normal distribution. \\
So
$c_{n-1,k}^{\ast}=c_{n-1,k}^{\ast\ast}=\frac{(2\pi)^{-\frac{n}{2}}}{\phi(\lambda_{k,\boldsymbol{q}})}$ and $c_{n-2,ij}^{\ast\ast}=\frac{(2\pi)^{-\frac{n}{2}}}{\phi(\lambda_{i,\boldsymbol{q}})\phi(\lambda_{j,\boldsymbol{q}})}$. Thus,
 \begin{align*}
&\mathbf{\Omega}_{\boldsymbol{q},ij}\\
&=\frac{2}{\overline{F}_{\mathbf{Z}}(\boldsymbol{\lambda_{q}})}\bigg\{\phi(\lambda_{i,\boldsymbol{q}})\phi(\lambda_{j,\boldsymbol{q}})\overline{E}^{\boldsymbol{\lambda}_{-ij,\boldsymbol{q}}}_{\mathbf{U}_{-ij}}[H(\boldsymbol{\gamma}^{T}\boldsymbol{\tau}_{ij,\boldsymbol{q}})]+\gamma_{j}\phi(\lambda_{i,\boldsymbol{q}})\overline{E}^{\boldsymbol{\lambda}_{-i,\boldsymbol{q}}}_{\mathbf{W}_{-i}}[H'(\boldsymbol{\gamma}^{T}\boldsymbol{\xi}_{i,\boldsymbol{q}})]\\
&~~~+\gamma_{i}\phi(\lambda_{j,\boldsymbol{q}})\overline{E}^{\boldsymbol{\lambda}_{-j,\boldsymbol{q}}}_{\mathbf{W}_{-j}}[H'(\boldsymbol{\gamma}^{T}\boldsymbol{\xi}_{j,\boldsymbol{q}})]+\gamma_{i}\gamma_{j}\overline{E}^{\boldsymbol{\lambda}_{\boldsymbol{q}}}_{\mathbf{X}}[H''(\boldsymbol{\gamma}^{T}\boldsymbol{X})]\bigg\}\\
&~~~-MTCE_{\mathbf{Z}}(\boldsymbol{\lambda_{q}})_{i}MTCE_{\mathbf{Z}}(\boldsymbol{\lambda_{q}})_{j},~i\neq j
\end{align*}
and
\begin{align*}
\mathbf{\Omega}_{\boldsymbol{q},ii}&=\frac{2}{\overline{F}_{\mathbf{Z}}(\boldsymbol{\lambda_{q}})}\bigg\{\lambda_{i,\boldsymbol{q}}\phi(\lambda_{i,\boldsymbol{q}})\overline{E}^{\boldsymbol{\lambda}_{-i,\boldsymbol{q}}}_{\boldsymbol{W}_{-i}}[H(\boldsymbol{\gamma}^{T}\boldsymbol{\xi}_{i,\boldsymbol{q}})]+\gamma_{i}\phi(\lambda_{i,\boldsymbol{q}})\overline{E}^{\boldsymbol{\lambda}_{-i,\boldsymbol{q}}}_{\mathbf{W}_{-i}}[H'(\boldsymbol{\gamma}^{T}\boldsymbol{\xi}_{i,\boldsymbol{q}})] \\ &+\gamma_{i}^{2}\overline{E}^{\boldsymbol{\lambda}_{\boldsymbol{q}}}_{\mathbf{X}}[H''(\boldsymbol{\gamma}^{T}\boldsymbol{X})]\bigg\}
+1-[MTCE_{\mathbf{Z}}(\boldsymbol{\lambda_{q}})_{i}]^{2},
\end{align*}
with
\begin{align*}
&MTCE_{\mathbf{Z}}(\boldsymbol{\lambda_{q}})_{k}=\frac{2}{\overline{F}_{\mathbf{Z}}(\boldsymbol{\lambda}_{\boldsymbol{q}})}\left\{\phi(\lambda_{k,\boldsymbol{q}})\overline{E}^{\boldsymbol{\lambda}_{-k,\boldsymbol{q}}}_{\boldsymbol{W}_{-k}}[H(\boldsymbol{\gamma}^{T}\boldsymbol{\xi}_{k,\boldsymbol{q}})]+\gamma_{k}\overline{E}^{\boldsymbol{\lambda_{q}}}_{\boldsymbol{X}}[H'(\boldsymbol{\gamma}^{T}\boldsymbol{X})]\right\},
\end{align*}
$~i,~j,~k=1,~2,\cdots,n,$
$\mathbf{Z}\sim GSN_{n}\left(\boldsymbol{0},~\boldsymbol{I_{n}},~\boldsymbol{\gamma},~H(\cdot)\right)$, $\mathbf{U}_{-ij}\sim N_{n-2}\left(\boldsymbol{0},~\boldsymbol{I_{n-2}}\right)$, $\mathbf{W}_{-k}\sim N_{n-1}\left(\boldsymbol{0},~\boldsymbol{I_{n-1}}\right)$, $\mathbf{X}\sim N_{n}(\boldsymbol{0},~\mathbf{I_{n}})$ and  $H'(\cdot)$ is the derivative of $H(\cdot)$.

 When $H(\cdot)=\Phi(\cdot)$(the cdf of $1$-dimensional standard normal distribution) in Example $4.1$, it will be an $n$-variate skew-normal distribution. Thus,
\begin{align*}
\mathbf{\Omega}_{\boldsymbol{q},ij}=&\frac{2}{\overline{F}_{\mathbf{Z}}(\boldsymbol{\lambda_{q}})}\bigg\{\phi(\lambda_{i,\boldsymbol{q}})\phi(\lambda_{j,\boldsymbol{q}})\overline{E}^{\boldsymbol{\lambda}_{-ij,\boldsymbol{q}}}_{\mathbf{U}_{-ij}}[\Phi(\boldsymbol{\gamma}^{T}\boldsymbol{\tau}_{ij,\boldsymbol{q}})]+\gamma_{j}\phi(\lambda_{i,\boldsymbol{q}})\overline{E}^{\boldsymbol{\lambda}_{-i,\boldsymbol{q}}}_{\mathbf{W}_{-i}}[\phi(\boldsymbol{\gamma}^{T}\boldsymbol{\xi}_{i,\boldsymbol{q}})]\\
&+\gamma_{i}\phi(\lambda_{j,\boldsymbol{q}})\overline{E}^{\boldsymbol{\lambda}_{-j,\boldsymbol{q}}}_{\mathbf{W}_{-j}}[\phi(\boldsymbol{\gamma}^{T}\boldsymbol{\xi}_{j,\boldsymbol{q}})]-\gamma_{i}\gamma_{j}\overline{E}^{\boldsymbol{\lambda}_{\boldsymbol{q}}}_{\mathbf{X}}[\boldsymbol{\gamma}^{T}\boldsymbol{X}\phi(\boldsymbol{\gamma}^{T}\boldsymbol{X})]\bigg\}\\
&-MTCE_{\mathbf{Z}}(\boldsymbol{\lambda_{q}})_{i}MTCE_{\mathbf{Z}}(\boldsymbol{\lambda_{q}})_{j},~i\neq j
\end{align*}
and
\begin{align*}
\mathbf{\Omega}_{\boldsymbol{q},ii}&=\frac{2}{\overline{F}_{\mathbf{Z}}(\boldsymbol{\lambda_{q}})}\bigg\{\lambda_{i,\boldsymbol{q}}\phi(\lambda_{i,\boldsymbol{q}})\overline{E}^{\boldsymbol{\lambda}_{-i,\boldsymbol{q}}}_{\boldsymbol{W}_{-i}}[\Phi(\boldsymbol{\gamma}^{T}\boldsymbol{\xi}_{i,\boldsymbol{q}})] +\gamma_{i}\phi(\lambda_{i,\boldsymbol{q}})\overline{E}^{\boldsymbol{\lambda}_{-i,\boldsymbol{q}}}_{\mathbf{W}_{-i}}[\phi(\boldsymbol{\gamma}^{T}\boldsymbol{\xi}_{i,\boldsymbol{q}})] \\ &-\gamma_{i}^{2}\overline{E}^{\boldsymbol{\lambda}_{\boldsymbol{q}}}_{\mathbf{X}}[\boldsymbol{\gamma}^{T}\boldsymbol{X}\phi(\boldsymbol{\gamma}^{T}\boldsymbol{X})]\bigg\}
+1-[MTCE_{\mathbf{Z}}(\boldsymbol{\lambda_{q}})_{i}]^{2},
\end{align*}
with
\begin{align*}
&MTCE_{\mathbf{Z}}(\boldsymbol{\lambda_{q}})_{k}=\frac{2}{\overline{F}_{\mathbf{Z}}(\boldsymbol{\lambda}_{\boldsymbol{q}})}\left\{\phi(\lambda_{k,\boldsymbol{q}})\overline{E}^{\boldsymbol{\lambda}_{-k,\boldsymbol{q}}}_{\boldsymbol{W}_{-k}}[\Phi(\boldsymbol{\gamma}^{T}\boldsymbol{\xi}_{k,\boldsymbol{q}})]+\gamma_{k}\overline{E}^{\boldsymbol{\lambda_{q}}}_{\boldsymbol{X}}[\phi(\boldsymbol{\gamma}^{T}\mathbf{X})]\right\},
\end{align*}
$~i,~j,~k=1,~2,\cdots,n,$
$\mathbf{Z}\sim SN_{n}\left(\boldsymbol{0},~\boldsymbol{I_{n}},~\boldsymbol{\gamma}\right)$, and $\mathbf{U}_{-ij}$, $\mathbf{W}_{-k}$ and $\mathbf{X}$ are same as those in Example 4.1. In addition, $\phi(\cdot)$ is the pdf of $1$-dimensional standard normal distribution.

 When $H(\cdot)=\frac{1}{2}$ in Example $4.1$, the MTCov for $n$-dimensional normal distribution is given,
\begin{align*}
&\mathbf{\Omega}_{\boldsymbol{q},ij}\\
&=\frac{1}{\overline{F}_{\mathbf{Z}}(\boldsymbol{\lambda_{q}})}\bigg\{\phi(\lambda_{i,\boldsymbol{q}})\phi(\lambda_{j,\boldsymbol{q}})\overline{F}_{\mathbf{U}_{-ij}}(\boldsymbol{\lambda}_{-ij,\boldsymbol{q}})\bigg\}-MTCE_{\mathbf{Z}}(\boldsymbol{\lambda_{q}})_{i}MTCE_{\mathbf{Z}}(\boldsymbol{\lambda_{q}})_{j},~i\neq j
\end{align*}
and
\begin{align*}
\mathbf{\Omega}_{\boldsymbol{q},ii}&=\frac{1}{\overline{F}_{\mathbf{Z}}(\boldsymbol{\lambda_{q}})}\bigg\{\lambda_{i,\boldsymbol{q}}\phi(\lambda_{i,\boldsymbol{q}})\overline{F}_{\boldsymbol{W}_{-i}}(\boldsymbol{\lambda}_{-i,\boldsymbol{q}})\bigg\}
+1-[MTCE_{\mathbf{Z}}(\boldsymbol{\lambda_{q}})_{i}]^{2},
\end{align*}
with
\begin{align*}
&MTCE_{\mathbf{Z}}(\boldsymbol{\lambda_{q}})_{k}=\frac{1}{\overline{F}_{\mathbf{Z}}(\boldsymbol{\lambda}_{\boldsymbol{q}})}\left\{\phi(\lambda_{k,\boldsymbol{q}})\overline{F}_{\boldsymbol{W}_{-k}}(\boldsymbol{\lambda}_{-k,\boldsymbol{q}})\right\},
\end{align*}
$~i,~j,~k=1,~2,\cdots,n,$
$\mathbf{Z}\sim N_{n}(\boldsymbol{0},~\mathbf{I_{n}})$, and $\mathbf{W}_{-k}$ and $\mathbf{U}_{-ij}$ are the same as those in Example $4.1$.\\
We observe that above result is coincide with the results of $(5.21)$ and $(5.22)$ in  Landsman et al. (2018).\\
$\mathbf{Example~4.2}$ (Generalized skew student-$t$ distribution). An n-dimensional generalized skew student-$t$ random vector $\mathbf{Y}$, with location parameter $\boldsymbol{\mu}$, scale matrix $\mathbf{\Sigma}$, $m>0$ degrees of freedom and skewing function $H(\cdot): \mathbb{R}\rightarrow\mathbb{R}$, has its density function as
\begin{align*}
f_{\boldsymbol{Y}}(\boldsymbol{y})=&\frac{2c_{n}}{\sqrt{|\mathbf{\Sigma}|}}\left[1+\frac{(\boldsymbol{y}-\boldsymbol{\mu})^{T}\mathbf{\Sigma}^{-1}(\boldsymbol{y}-\boldsymbol{\mu})}{m}\right]^{-\frac{m+n}{2}}H\left(\boldsymbol{\gamma}^{T}\mathbf{\Sigma}^{-\frac{1}{2}}(\boldsymbol{y}-\boldsymbol{\mu})\right),~\boldsymbol{y}\in\mathbb{R}^{n},
 \end{align*}
 where $\boldsymbol{\gamma}=(\gamma_{1},~\gamma_{2},~\cdots,~\gamma_{n})^{T}\in\mathbb{R}^{n}$ and $c_{n}=\frac{\Gamma\left(\frac{m+n}{2}\right)}{\Gamma(m/2)(m\pi)^{\frac{n}{2}}}$. We
denote it by $\mathbf{Y}\sim GSSt_{n}(\boldsymbol{\mu},~\mathbf{\Sigma},~\boldsymbol{\gamma},~m,~H(\cdot))$. The density generator in this case is $$g_{n}(u)=\left(1+\frac{2u}{m}\right)^{-(m+n)/2},$$~and so $\overline{G}_{n}(t)$ and $\overline{\mathcal{G}}_{n}(t)$ can be expressed, respectively, as
 $$\overline{G}_{n}(t)=\frac{m}{m+n-2}\left(1+\frac{2t}{m}\right)^{-(m+n-2)/2}$$
  and
 $$\overline{\mathcal{G}}_{n}(t)=\frac{m}{m+n-2}\frac{m}{m+n-4}\left(1+\frac{2t}{m}\right)^{-(m+n-4)/2}.$$ In addition, \begin{align*}
 c_{n}^{\ast}&=\frac{(m+n-2)\Gamma(n/2)}{(2\pi)^{n/2}m}\left[\int_{0}^{\infty}t^{n/2-1}\left(1+\frac{2t}{m}\right)^{-(m+n-2)/2}\mathrm{d}t\right]^{-1}\\
 &=\frac{(m+n-2)\Gamma(n/2)}{(m\pi)^{n/2}mB(\frac{n}{2},~\frac{m-2}{2})},~if~m>2
 \end{align*} and
 \begin{align*}
 c_{n}^{\ast\ast}&=\frac{(m+n-2)(m+n-4)\Gamma(n/2)}{(2\pi)^{n/2}m^{2}}\left[\int_{0}^{\infty}t^{n/2-1}\left(1+\frac{2t}{m}\right)^{-(m+n-4)/2}\mathrm{d}t\right]^{-1}\\
 &=\frac{(m+n-2)(m+n-4)\Gamma(n/2)}{(m\pi)^{n/2}m^{2}B(\frac{n}{2},~\frac{m-4}{2})},~if~m>4,
 \end{align*}
 where $\Gamma(\cdot)$ and $B(\cdot,\cdot)$ are Gamma function and Beta function, respectively. And~ $$H\left(\mathbf{\Sigma}^{-\frac{1}{2}}(\boldsymbol{y}-\boldsymbol{\mu})\right)=H\left(\boldsymbol{\gamma}^{T}\mathbf{\Sigma}^{-\frac{1}{2}}(\boldsymbol{y}-\boldsymbol{\mu})\right).$$
  Since
  \begin{align*}
  f_{\mathbf{W}_{-k}}(\boldsymbol{w})&=c_{n-1,k}^{\ast}\frac{m}{m+n-2}\left(1+\frac{\lambda_{k,\boldsymbol{q}}^{2}}{m}\right)^{-\frac{m+n-2}{2}}\left(1+\frac{\boldsymbol{w}^{T}\Delta_{k}^{-1}\boldsymbol{w}}{m-1}\right)^{-\frac{m+n-2}{2}}\\
 &=St_{n-1}(\boldsymbol{0},~\Delta_{k},~m-1),~\boldsymbol{w}\in \mathbb{R}^{n-1},
 \end{align*}
 \begin{align*}
 f_{\mathbf{V}_{-k}}(\boldsymbol{v})&=c_{n-1,k}^{\ast\ast}\frac{m}{m+n-2}\frac{m}{m+n-4}\left(1+\frac{\lambda_{k,\boldsymbol{q}}^{2}}{m}\right)^{-\frac{m+n-4}{2}}\left(1+\frac{\boldsymbol{v}^{T}\Lambda_{k}^{-1}\boldsymbol{v}}{m-3}\right)^{-\frac{m+n-4}{2}}\\
 &=St_{n-1}(\boldsymbol{0},~\Lambda_{k},~m-3),~\boldsymbol{v}\in \mathbb{R}^{n-1}
 \end{align*}
 and
\begin{align*}
&f_{\mathbf{U}_{-ij}}(\boldsymbol{u})\\
&=c_{n-2,ij}^{\ast\ast}\frac{m}{m+n-2}\frac{m}{m+n-4}\left(1+\frac{\lambda_{i,\boldsymbol{q}}^{2}+\lambda_{j,\boldsymbol{q}}^{2}}{m}\right)^{-\frac{m+n-4}{2}}\left(1+\frac{\boldsymbol{u}^{T}\Theta_{ij}^{-1}\boldsymbol{u}}{m-2}\right)^{-\frac{m+n-4}{2}}\\
 &=St_{n-2}(\boldsymbol{0},~\Theta_{ij},~m-2),~\boldsymbol{u}\in \mathbb{R}^{n-2},
 \end{align*}
so that
$$c_{n-1,k}^{\ast}=\frac{\Gamma\left(\frac{m+n-2}{2}\right)(m+n-2)}{\Gamma\left(\frac{m-1}{2}\right)\pi^{\frac{n-1}{2}}(m-1)^{\frac{n-1}{2}}m^{\frac{n+m}{2}}}\left(m+\lambda_{k,\boldsymbol{q}}^{2}\right)^{\frac{m+n-2}{2}},$$
$$c_{n-1,k}^{\ast\ast}=\frac{\Gamma\left(\frac{m+n-4}{2}\right)(m+n-2)(m+n-4)}{\Gamma\left(\frac{m-3}{2}\right)\pi^{\frac{n-1}{2}}m^{\frac{m+n}{2}}}\left(m+\lambda_{k,\boldsymbol{q}}^{2}\right)^{\frac{m-3}{2}}$$
 and
 $$c_{n-2,ij}^{\ast\ast}=\frac{\Gamma\left(\frac{m+n-4}{2}\right)(m+n-2)(m+n-4)}{\Gamma\left(\frac{m-2}{2}\right)\pi^{\frac{n-2}{2}}m^{\frac{m+n}{2}}}\left(m+\lambda_{i,\boldsymbol{q}}^{2}+\lambda_{j,\boldsymbol{q}}^{2}\right)^{\frac{m-2}{2}}.$$
Then
 \begin{align*}
&\mathbf{\Omega}_{\boldsymbol{q},ij}=\frac{2}{\overline{F}_{\mathbf{Z}}(\boldsymbol{\lambda_{q}})}\bigg\{\frac{c_{n}}{c_{n-2,ij}^{\ast\ast}}\overline{E}^{\boldsymbol{\lambda}_{-ij,\boldsymbol{q}}}_{\mathbf{U}_{-ij}}[H(\boldsymbol{\gamma}^{T}\boldsymbol{\tau}_{ij,\boldsymbol{q}})]+\frac{c_{n}}{c_{n-1,i}^{\ast\ast}}\gamma_{j}\overline{E}^{\boldsymbol{\lambda}_{-i,\boldsymbol{q}}}_{\mathbf{V}_{-i}}[H'(\boldsymbol{\gamma}^{T}\boldsymbol{\eta}_{i,\boldsymbol{q}})]\\
&+\frac{c_{n}}{c_{n-1,j}^{\ast\ast}}\gamma_{i}\overline{E}^{\boldsymbol{\lambda}_{-j,\boldsymbol{q}}}_{\mathbf{V}_{-j}}[H'(\boldsymbol{\gamma}^{T}\boldsymbol{\eta}_{j,\boldsymbol{q}})]+\frac{m^{2}}{(m-2)(m-4)}\gamma_{i}\gamma_{j}\overline{E}^{\boldsymbol{\lambda}_{\boldsymbol{q}}}_{\mathbf{X}^{\ast\ast}}[H''(\boldsymbol{\gamma}^{T}\boldsymbol{X}^{\ast\ast})]\bigg\}\\
&-MTCE_{\mathbf{Z}}(\boldsymbol{\lambda_{q}})_{i}MTCE_{\mathbf{Z}}(\boldsymbol{\lambda_{q}})_{j},~i\neq j,~m>4
\end{align*}
and
\begin{align*}
&\mathbf{\Omega}_{\boldsymbol{q},ii}
=\frac{2}{\overline{F}_{\mathbf{Z}}(\boldsymbol{\lambda_{q}})}\bigg\{\frac{c_{n}}{c_{n-1,i}^{\ast}}\lambda_{i,\boldsymbol{q}}\overline{E}^{\boldsymbol{\lambda}_{-i,\boldsymbol{q}}}_{\boldsymbol{W}_{-i}}[H(\boldsymbol{\gamma}^{T}\boldsymbol{\xi}_{i,\boldsymbol{q}})] +\frac{c_{n}}{c_{n-1,i}^{\ast\ast}}\gamma_{i}\overline{E}^{\boldsymbol{\lambda}_{-i,\boldsymbol{q}}}_{\mathbf{V}_{-i}}[H'(\boldsymbol{\gamma}^{T}\boldsymbol{\eta}_{i,\boldsymbol{q}})] \\ &~~~+\frac{m^{2}}{(m-2)(m-4)}\gamma_{i}^{2}\overline{E}^{\boldsymbol{\lambda}_{\boldsymbol{q}}}_{\mathbf{X}^{\ast\ast}}[H''(\boldsymbol{\gamma}^{T}\boldsymbol{X}^{\ast\ast})]\bigg\}
+\frac{m}{m-2}\frac{\overline{F}_{\mathbf{Z^{\ast}}}(\boldsymbol{\lambda_{q}})}{\overline{F}_{\mathbf{Z}}(\boldsymbol{\lambda_{q}})}\\
&~~~-[MTCE_{\mathbf{Z}}(\boldsymbol{\lambda_{q}})_{i}]^{2},~m>4,
\end{align*}
with
\begin{align*}
MTCE_{\mathbf{Z}}(\boldsymbol{\lambda_{q}})_{k}=&\frac{1}{\overline{F}_{\mathbf{Z}}(\boldsymbol{\lambda}_{\boldsymbol{q}})}\bigg\{\frac{c_{n}}{c_{n-1,k}^{\ast}}\overline{E}^{\boldsymbol{\lambda}_{-k,\boldsymbol{q}}}_{\boldsymbol{W}_{-k}}[H(\boldsymbol{\gamma}^{T}\boldsymbol{\xi}_{k,\boldsymbol{q}})]\\
&+\frac{2m\gamma_{k}}{m+n-2}\overline{E}^{\boldsymbol{\lambda_{q}}}_{\boldsymbol{M}}\left[\left(1+\frac{\boldsymbol{M}^{ T}\boldsymbol{M}^{\ast}}{m}\right)H'(\boldsymbol{\gamma^{T}\boldsymbol{M}})\right]\bigg\},
\end{align*}
 $i,~j,~k=1,~2,\cdots,n$, $\mathbf{Z}\sim GSSt_{n}(\boldsymbol{0},~\mathbf{I_{n}},~\boldsymbol{\gamma},~m,~H(\cdot)),$ \\$\mathbf{Z}^{\ast}\sim GSE_{n}\left(\boldsymbol{0},~\boldsymbol{I_{n}},~\boldsymbol{\gamma},~m,~\overline{G}_{n},~H(\cdot)\right)$, $\mathbf{X}^{\ast\ast}\sim E_{n}(\boldsymbol{0},~\mathbf{I_{n}},~\overline{\mathcal{G}}_{n})$, \\$\mathbf{U}_{-ij}\sim St_{n-2}\left(\boldsymbol{0},~\Theta_{ij},~m-2\right)$, $\mathbf{V}_{-k}\sim St_{n-1}\left(\boldsymbol{0},~\Lambda_{k},~m-3\right)$, \\$\mathbf{W}_{-k}\sim St_{n-1}(\boldsymbol{0},~\mathbf{\bigtriangleup}_{k},~m-1),$
$\bigtriangleup_{k}=\left(\frac{m+\lambda_{k,q}^{2}}{m-1}\right)\mathbf{I}_{n-1} $, $\Lambda_{k}=\left(\frac{m+\lambda_{k,q}^{2}}{m-3}\right)\mathbf{I}_{n-1}$, $\Theta_{ij}=\left(\frac{m+\lambda_{i,\boldsymbol{q}}^{2}+\lambda_{j,\boldsymbol{q}}^{2}}{m-2}\right)\mathbf{I}_{n-2}$ and $\mathbf{M}\sim St_{n}(\boldsymbol{0},~\mathbf{I_{n}},m)$~(student-$t$ distribution).\\
Therefore, we can simplify them further as follows:
$$\frac{c_{n}}{c_{n-1,k}^{\ast}}=\frac{\Gamma(\frac{m-1}{2})\left(\frac{m-1}{m}\right)^{(n-1)/2}}{\Gamma(\frac{m}{2})\sqrt{\pi/m}}\left(1+\frac{\lambda_{k,\boldsymbol{q}}^{2}}{m}\right)^{-(m+n-2)/2},$$
$$\frac{c_{n}}{c_{n-1,k}^{\ast\ast}}=\frac{\Gamma(\frac{m-3}{2})(m+n)m^{\frac{m}{2}}}{4\Gamma(\frac{m}{2})(m+n-4)\sqrt{\pi}}\left(m+\lambda_{k,\boldsymbol{q}}^{2}\right)^{-(m-3)/2}$$
and
$$\frac{c_{n}}{c_{n-2,ij}^{\ast\ast}}=\frac{(m+n)m^{\frac{m-2}{2}}}{2(m+n-4)\pi}\left(m+\lambda_{i,\boldsymbol{q}}^{2}+\lambda_{j,\boldsymbol{q}}^{2}\right)^{-(m-2)/2}.$$

 When $H(\cdot)=T(\cdot)$(the cdf of $1$-dimensional standard student-$t$ distribution) in Example $4.2$, it is an $n$-dimensional skew student-$t$ distribution. Thus,
 \begin{align*}
&\mathbf{\Omega}_{\boldsymbol{q},ij}=\frac{2}{\overline{F}_{\mathbf{Z}}(\boldsymbol{\lambda_{q}})}\bigg\{\frac{c_{n}}{c_{n-2,ij}^{\ast\ast}}\overline{E}^{\boldsymbol{\lambda}_{-ij,\boldsymbol{q}}}_{\mathbf{U}_{-ij}}[T(\boldsymbol{\gamma}^{T}\boldsymbol{\tau}_{ij,\boldsymbol{q}})]+\frac{c_{n}}{c_{n-1,i}^{\ast\ast}}\gamma_{j}\overline{E}^{\boldsymbol{\lambda}_{-i,\boldsymbol{q}}}_{\mathbf{V}_{-i}}[\boldsymbol{t}(\boldsymbol{\gamma}^{T}\boldsymbol{\eta}_{i,\boldsymbol{q}})]\\
&-\frac{(m+1)m}{(m-2)(m-4)}\gamma_{i}\gamma_{j}\overline{E}^{\boldsymbol{\lambda}_{\boldsymbol{q}}}_{\mathbf{X}^{\ast\ast}}\left[\left(1+\frac{(\boldsymbol{\gamma}^{T}\boldsymbol{X}^{\ast\ast})^{2}}{m}\right)^{-1}\boldsymbol{\gamma}^{T}\boldsymbol{X}^{\ast\ast}\boldsymbol{t}(\boldsymbol{\gamma}^{T}\boldsymbol{X}^{\ast\ast})\right]\\
&+\frac{c_{n}}{c_{n-1,j}^{\ast\ast}}\gamma_{i}\overline{E}^{\boldsymbol{\lambda}_{-j,\boldsymbol{q}}}_{\mathbf{V}_{-j}}[\boldsymbol{t}(\boldsymbol{\gamma}^{T}\boldsymbol{\eta}_{j,\boldsymbol{q}})]\bigg\}-MTCE_{\mathbf{Z}}(\boldsymbol{\lambda_{q}})_{i}MTCE_{\mathbf{Z}}(\boldsymbol{\lambda_{q}})_{j},~i\neq j,~m>4
\end{align*}
and
\begin{align*}
&\mathbf{\Omega}_{\boldsymbol{q},ii}
=\frac{2}{\overline{F}_{\mathbf{Z}}(\boldsymbol{\lambda_{q}})}\bigg\{\frac{c_{n}}{c_{n-1,i}^{\ast}}\lambda_{i,\boldsymbol{q}}\overline{E}^{\boldsymbol{\lambda}_{-i,\boldsymbol{q}}}_{\boldsymbol{W}_{-i}}[T(\boldsymbol{\gamma}^{T}\boldsymbol{\xi}_{i,\boldsymbol{q}})] +\frac{c_{n}}{c_{n-1,i}^{\ast\ast}}\gamma_{i}\overline{E}^{\boldsymbol{\lambda}_{-i,\boldsymbol{q}}}_{\mathbf{V}_{-i}}[\boldsymbol{t}(\boldsymbol{\gamma}^{T}\boldsymbol{\eta}_{i,\boldsymbol{q}})] \\ &-\frac{(m+1)m}{(m-2)(m-4)}\gamma_{i}^{2}\overline{E}^{\boldsymbol{\lambda}_{\boldsymbol{q}}}_{\mathbf{X}^{\ast\ast}}\left[\left(1+\frac{(\boldsymbol{\gamma}^{T}\boldsymbol{X}^{\ast\ast})^{2}}{m}\right)^{-1}\boldsymbol{\gamma}^{T}\boldsymbol{X}^{\ast\ast}\boldsymbol{t}(\boldsymbol{\gamma}^{T}\boldsymbol{X}^{\ast\ast})\right]\bigg\}\\
&+\frac{m}{m-2}\frac{\overline{F}_{\mathbf{Z^{\ast}}}(\boldsymbol{\lambda_{q}})}{\overline{F}_{\mathbf{Z}}(\boldsymbol{\lambda_{q}})}-[MTCE_{\mathbf{Z}}(\boldsymbol{\lambda_{q}})_{i}]^{2},~m>4,
\end{align*}
with
\begin{align*}
MTCE_{\mathbf{Z}}(\boldsymbol{\lambda_{q}})_{k}=&\frac{1}{\overline{F}_{\mathbf{Z}}(\boldsymbol{\lambda}_{\boldsymbol{q}})}\bigg\{\frac{c_{n}}{c_{n-1,k}^{\ast}}\overline{E}^{\boldsymbol{\lambda}_{-k,\boldsymbol{q}}}_{\boldsymbol{W}_{-k}}[T(\boldsymbol{\gamma}^{T}\boldsymbol{\xi}_{k,\boldsymbol{q}})]\\
&+\frac{2m\gamma_{k}}{m+n-2}\overline{E}^{\boldsymbol{\lambda_{q}}}_{\boldsymbol{M}}\left[\left(1+\frac{\boldsymbol{M}^{ T}\boldsymbol{M}}{m}\right)\boldsymbol{t}(\boldsymbol{\gamma^{T}\boldsymbol{M}})\right]\bigg\},
\end{align*}
 $i,~j,~k=1,~2,\cdots,n$,
$\mathbf{Z}\sim SSt_{n}(\boldsymbol{0},~\mathbf{I_{n}},~\boldsymbol{\gamma},~m),$\\ $\mathbf{Z}^{\ast}\sim GSE_{n}\left(\boldsymbol{0},~\boldsymbol{I_{n}},~\boldsymbol{\gamma},~m,~\overline{G}_{n},~T(\cdot)\right)$,
and $\boldsymbol{t}(\cdot)$ is the pdf of $1$-dimensional standard student-$t$ distribution. Moreover, $\mathbf{M}$, $\mathbf{X}^{\ast\ast}$, $\mathbf{W}_{-k}$, $\mathbf{V}_{-k}$ and $\mathbf{U}_{-ij}$ are the same as those in Example $4.2$.

When $H(\cdot)=\frac{1}{2}$ in Example $4.2$, we obtain a formula of MTCov for  student-$t$ distribution in Landsman et al. (2018):
\begin{align*}
\mathbf{\Omega}_{\boldsymbol{q},ij}=&\frac{1}{\overline{F}_{\mathbf{Z}}(\boldsymbol{\lambda_{q}})}\bigg\{\frac{c_{n}}{c_{n-2,ij}^{\ast\ast}}\overline{F}_{\mathbf{U}_{-ij}}(\boldsymbol{\lambda}_{-ij,\boldsymbol{q}})\bigg\}-MTCE_{\mathbf{Z}}(\boldsymbol{\lambda_{q}})_{i}MTCE_{\mathbf{Z}}(\boldsymbol{\lambda_{q}})_{j},~i\neq j
\end{align*}
and
\begin{align*}
&\mathbf{\Omega}_{\boldsymbol{q},ii}\\
&=\frac{1}{\overline{F}_{\mathbf{Z}}(\boldsymbol{\lambda_{q}})}\bigg\{\frac{c_{n}}{c_{n-1,i}^{\ast}}\lambda_{i,\boldsymbol{q}}\overline{F}_{\boldsymbol{W}_{-i}}(\boldsymbol{\lambda}_{-i,\boldsymbol{q}}) \bigg\}
+\frac{m}{m-2}\frac{\overline{F}_{\mathbf{Z^{\ast}}}(\boldsymbol{\lambda_{q}})}{\overline{F}_{\mathbf{Z}}(\boldsymbol{\lambda_{q}})}-[MTCE_{\mathbf{Z}}(\boldsymbol{\lambda_{q}})_{i}]^{2},
\end{align*}
$m>2,$ with
\begin{align*}
MTCE_{\mathbf{Z}}(\boldsymbol{\lambda_{q}})_{k}=&\frac{1}{\overline{F}_{\mathbf{Z}}(\boldsymbol{\lambda}_{\boldsymbol{q}})}\bigg\{\frac{c_{n}}{c_{n-1,k}^{\ast}}\overline{F}_{\boldsymbol{W}_{-k}}(\boldsymbol{\lambda}_{-k,\boldsymbol{q}})\bigg\},
\end{align*}
 $i,~j,~k=1,~2,\cdots,n$, $\mathbf{Z}\sim St_{n}(\boldsymbol{0},~\mathbf{I_{n}},~m)$ and $\mathbf{Z}^{\ast}\sim E_{n}\left(\boldsymbol{0},~\boldsymbol{I_{n}},~m,~\overline{G}_{n}\right)$. In addition, $\mathbf{U}_{-ij}$ and $\mathbf{W}_{-k}$ are the same as those in Example $4.2$.\\
$\mathbf{Example~4.3}$ (Generalized skew-logistic distribution). An $n$-variate generalized skew-logistic random vector $\mathbf{Y}$, with location parameter $\boldsymbol{\mu}$, scale matrix $\mathbf{\Sigma}$ and skewing function $H(\cdot): \mathbb{R}\rightarrow\mathbb{R}$, has its density function as
 \begin{align*}
f_{\boldsymbol{Y}}(\boldsymbol{y})=&\frac{2c_{n}}{\sqrt{|\mathbf{\Sigma}|}}\frac{\exp\left\{-\frac{1}{2}(\boldsymbol{y}-\boldsymbol{\mu})^{T}\mathbf{\Sigma}^{-1}(\boldsymbol{y}-\boldsymbol{\mu})\right\}}{\left[1+\exp\left\{-\frac{1}{2}(\boldsymbol{y}-\boldsymbol{\mu})^{T}\mathbf{\Sigma}^{-1}(\boldsymbol{y}-\boldsymbol{\mu})\right\}\right]^{2}}H\left(\boldsymbol{\gamma}^{T}\mathbf{\Sigma}^{-\frac{1}{2}}(\boldsymbol{y}-\boldsymbol{\mu})\right),
 \end{align*}
 $~\boldsymbol{y}\in\mathbb{R}^{n},$
 where $\boldsymbol{\gamma}=(\gamma_{1},~\gamma_{2},~\cdots,~\gamma_{n})^{T}$,
\begin{align*}
c_{n}=\frac{1}{(2\pi)^{n/2}\Psi_{2}^{\ast}(-1,\frac{n}{2},1)}.
\end{align*}
We denote it by $\mathbf{Y}\sim GSLo_{n}(\boldsymbol{\mu},~\mathbf{\Sigma},~\boldsymbol{\gamma},~H(\cdot))$. The density generator in this case is
 $$g_{n}(u)=\frac{\exp\{-u\}}{[1+\exp\{-u\}]^{2}},$$and  $\overline{G}_{n}(t)$ and $\overline{\mathcal{G}}_{n}(t)$ are given by
 $$\overline{G}_{n}(t)=\frac{\exp(-t)}{1+\exp(-t)},\;\;
 \overline{\mathcal{G}}_{n}(t)=\ln\left[1+\exp(-t)\right].$$ In addition,
 \begin{align*}
 c_{n}^{\ast}=\frac{1}{(2\pi)^{n/2}\Psi_{1}^{\ast}(-1,\frac{n}{2},1)}
\end{align*}
 and
 \begin{align*}
 c_{n}^{\ast\ast}=\frac{1}{(2\pi)^{n/2}\Psi_{1}^{\ast}(-1,\frac{n}{2}+1,1)}.
\end{align*} And $H\left(\mathbf{\Sigma}^{-\frac{1}{2}}(\boldsymbol{y}-\boldsymbol{\mu})\right)=H\left(\boldsymbol{\gamma}^{T}\mathbf{\Sigma}^{-\frac{1}{2}}(\boldsymbol{y}-\boldsymbol{\mu})\right)$.
 Since
 $$f_{\mathbf{W}_{-k}}(\boldsymbol{w})=c_{n-1,k}^{\ast}\frac{\exp\left(-\frac{1}{2}\boldsymbol{w}^{T}\boldsymbol{w}-\frac{1}{2}\lambda_{k,\boldsymbol{q}}^{2}\right)}{\left[1+\exp\left(-\frac{1}{2}\boldsymbol{w}^{T}\boldsymbol{w}-\frac{1}{2}\lambda_{k,\boldsymbol{q}}^{2}\right)\right]^{2}},~\boldsymbol{w}\in \mathbb{R}^{n-1},$$
 $$f_{\mathbf{V}_{-k}}(\boldsymbol{v})=c_{n-1,k}^{\ast\ast}\ln\left[1+\exp\left(-\frac{1}{2}\boldsymbol{v}^{T}\boldsymbol{v}-\frac{1}{2}\lambda_{k,\boldsymbol{q}}^{2}\right)\right],~\boldsymbol{v}\in \mathbb{R}^{n-1}$$
 and
$$f_{\mathbf{U}_{-ij}}(\boldsymbol{u})=c_{n-2,ij}^{\ast\ast}\ln\left[1+\exp\left(-\frac{1}{2}\boldsymbol{u}^{T}\boldsymbol{u}-\frac{1}{2}\lambda_{i,\boldsymbol{q}}^{2}-\frac{1}{2}\lambda_{j,\boldsymbol{q}}^{2}\right)\right],~\boldsymbol{u}\in \mathbb{R}^{n-2},$$
so that
\begin{align*}
c_{n-1,k}^{\ast}&=\frac{\Gamma((n-1)/2)\exp\{\frac{\lambda_{k,\boldsymbol{q}}^{2}}{2}\}}{(2\pi)^{(n-1)/2}}\left[\int_{0}^{\infty}\frac{t^{(n-3)/2}\exp\{-t\}}{1+\exp\{-\frac{\lambda_{k,\boldsymbol{q}}^{2}}{2}\}\exp\{-t\}}\mathrm{d}t\right]^{-1}\\
&=\frac{\exp\{\frac{\lambda_{k,\boldsymbol{q}}^{2}}{2}\}}{(2\pi)^{(n-1)/2}\Psi_{1}^{\ast}(-\exp\{-\frac{\lambda_{k,\boldsymbol{q}}^{2}}{2}\},~\frac{n-1}{2},~1)},
\end{align*}
\begin{align*}
c_{n-1,k}^{\ast\ast}&=\frac{\Gamma((n-1)/2)}{(2\pi)^{(n-1)/2}}\left\{\int_{0}^{\infty}t^{(n-3)/2}\ln\left[1+\exp\left(-\frac{1}{2}\lambda_{k,\boldsymbol{q}}^{2}\right)\exp(-t)\right]\mathrm{d}t\right\}^{-1},
\end{align*}
and
\begin{align*}
&c_{n-2,ij}^{\ast\ast}\\
&=\frac{\Gamma((n-2)/2)}{(2\pi)^{(n-2)/2}}\left\{\int_{0}^{\infty}t^{(n-4)/2}\ln\left[1+\exp\left(-\frac{1}{2}\lambda_{i,\boldsymbol{q}}^{2}-\frac{1}{2}\lambda_{j,\boldsymbol{q}}^{2}\right)\exp(-t)\right]\mathrm{d}t\right\}^{-1}.
\end{align*}
Then
 \begin{align*}
&\mathbf{\Omega}_{\boldsymbol{q},ij}=\frac{2}{\overline{F}_{\mathbf{Z}}(\boldsymbol{\lambda_{q}})}\bigg\{\frac{c_{n}}{c_{n-2,ij}^{\ast\ast}}\overline{E}^{\boldsymbol{\lambda}_{-ij,\boldsymbol{q}}}_{\mathbf{U}_{-ij}}[H(\boldsymbol{\gamma}^{T}\boldsymbol{\tau}_{ij,\boldsymbol{q}})]+\frac{c_{n}}{c_{n-1,i}^{\ast\ast}}\gamma_{j}\overline{E}^{\boldsymbol{\lambda}_{-i,\boldsymbol{q}}}_{\mathbf{V}_{-i}}[H'(\boldsymbol{\gamma}^{T}\boldsymbol{\eta}_{i,\boldsymbol{q}})]\\
&+\frac{c_{n}}{c_{n-1,j}^{\ast\ast}}\gamma_{i}\overline{E}^{\boldsymbol{\lambda}_{-j,\boldsymbol{q}}}_{\mathbf{V}_{-j}}[H'(\boldsymbol{\gamma}^{T}\boldsymbol{\eta}_{j,\boldsymbol{q}})]+\frac{\Psi_{1}^{\ast}(-1,\frac{n}{2}+1,1)}{\Psi_{2}^{\ast}(-1,\frac{n}{2},1)}\gamma_{i}\gamma_{j}\overline{E}^{\boldsymbol{\lambda}_{\boldsymbol{q}}}_{\mathbf{X}^{\ast\ast}}[H''(\boldsymbol{\gamma}^{T}\boldsymbol{X}^{\ast\ast})]\bigg\}\\
&-MTCE_{\mathbf{Z}}(\boldsymbol{\lambda_{q}})_{i}MTCE_{\mathbf{Z}}(\boldsymbol{\lambda_{q}})_{j},~i\neq j,
\end{align*}
and
\begin{align*}
\mathbf{\Omega}_{\boldsymbol{q},ii}
&=\frac{2}{\overline{F}_{\mathbf{Z}}(\boldsymbol{\lambda_{q}})}\bigg\{\frac{c_{n}}{c_{n-1,i}^{\ast}}\lambda_{i,\boldsymbol{q}}\overline{E}^{\boldsymbol{\lambda}_{-i,\boldsymbol{q}}}_{\boldsymbol{W}_{-i}}[H(\boldsymbol{\gamma}^{T}\boldsymbol{\xi}_{i,\boldsymbol{q}})] +\frac{c_{n}}{c_{n-1,i}^{\ast\ast}}\gamma_{i}\overline{E}^{\boldsymbol{\lambda}_{-i,\boldsymbol{q}}}_{\mathbf{V}_{-i}}[H'(\boldsymbol{\gamma}^{T}\boldsymbol{\eta}_{i,\boldsymbol{q}})] \\ &+\frac{\Psi_{1}^{\ast}(-1,\frac{n}{2}+1,1)}{\Psi_{2}^{\ast}(-1,\frac{n}{2},1)}\gamma_{i}^{2}\overline{E}^{\boldsymbol{\lambda}_{\boldsymbol{q}}}_{\mathbf{X}^{\ast\ast}}[H''(\boldsymbol{\gamma}^{T}\boldsymbol{X}^{\ast\ast})]\bigg\}
+\frac{\Psi_{1}^{\ast}(-1,\frac{n}{2},1)}{\Psi_{2}^{\ast}(-1,\frac{n}{2},1)}\frac{\overline{F}_{\mathbf{Z^{\ast}}}(\boldsymbol{\lambda_{q}})}{\overline{F}_{\mathbf{Z}}(\boldsymbol{\lambda_{q}})}\\
&-[MTCE_{\mathbf{Z}}(\boldsymbol{\lambda_{q}})_{i}]^{2},
\end{align*}
with
\begin{align*} MTCE_{\mathbf{Z}}(\boldsymbol{\lambda_{q}})_{k}&=\frac{2}{\overline{F}_{\mathbf{Z}}(\boldsymbol{\lambda}_{\boldsymbol{q}})}\bigg\{\frac{c_{n}}{c_{n-1,k}^{\ast}}\overline{E}^{\boldsymbol{\lambda}_{-k,\boldsymbol{q}}}_{\boldsymbol{W}_{-k}}[H(\boldsymbol{\gamma}^{T}\boldsymbol{\xi}_{k,\boldsymbol{q}})]\\
&+2\gamma_{k}\overline{E}^{\boldsymbol{\lambda_{q}}}_{\boldsymbol{M}}\left[\left(1+\exp\left\{-\frac{\boldsymbol{M}^{ T}\boldsymbol{M}}{2}\right\}\right)H'(\boldsymbol{\gamma^{T}\boldsymbol{M}})\right]\bigg\},
\end{align*}
 $i,~j,~k=1,~2,\cdots,n$, $\mathbf{Z}\sim GSLo_{n}(\boldsymbol{0},~\mathbf{I_{n}},~\boldsymbol{\gamma},~H(\cdot)),$ \\$\mathbf{Z}^{\ast}\sim GSE_{n}\left(\boldsymbol{0},~\boldsymbol{I_{n}},~\boldsymbol{\gamma},~\overline{G}_{n},~H(\cdot)\right)$, $\mathbf{X}^{\ast\ast}\sim E_{n}(\boldsymbol{0},~\mathbf{I_{n}},~\overline{\mathcal{G}}_{n})$ and $\mathbf{M}\sim Lo_{n}(\boldsymbol{0},~\mathbf{I_{n}})$ (logistic distribution).
In addition, $\Psi_{\mu}^{\ast}(z,s,a)$ is the generalized Hurwitz-Lerch zeta function defined by (cf.
Lin et al.(2006))
$$\Psi_{\mu}^{\ast}(z,s,a)=\frac{1}{\Gamma(\mu)}\sum_{n=0}^{\infty}\frac{\Gamma(\mu+n)}{n!}\frac{z^{n}}{(n+a)^{s}},$$
which has an integral representation
$$\Psi_{\mu}^{\ast}(z,s,a)=\frac{1}{\Gamma(s)}\int_{0}^{\infty}\frac{t^{s-1}e^{-at}}{(1-ze^{-t})^{\mu}}\mathrm{d}t,$$
where $\mathcal{R}(a)>0$; $\mathcal{R}(s)>0$ when $|z|\leq1~(z\neq1)$; $\mathcal{R}(s)>1$ when $z=1$.\\
Therefore, $$\frac{c_{n}}{c_{n-1,k}^{\ast}}=\frac{\Psi_{1}^{\ast}\left(-\exp\{-\frac{\lambda_{k,\boldsymbol{q}}^{2}}{2}\},~\frac{n-1}{2},~1\right)\phi(\lambda_{k,\boldsymbol{q}})}{\Psi_{2}^{\ast}(-1,\frac{n}{2},1)},$$ $$\frac{c_{n}}{c_{n-1,k}^{\ast\ast}}=\frac{\int_{0}^{\infty}t^{(n-3)/2}\ln\left[1+\exp\left(-\frac{1}{2}\lambda_{k,\boldsymbol{q}}^{2}\right)\exp(-t)\right]\mathrm{d}t}{\Gamma((n-1)/2)\sqrt{2\pi}\Psi_{2}^{\ast}(-1,\frac{n}{2},1)}$$
and
$$\frac{c_{n}}{c_{n-2,ij}^{\ast\ast}}=\frac{\int_{0}^{\infty}t^{(n-4)/2}\ln\left[1+\exp\left(-\frac{1}{2}\lambda_{i,\boldsymbol{q}}^{2}-\frac{1}{2}\lambda_{j,\boldsymbol{q}}^{2}\right)\exp(-t)\right]\mathrm{d}t}{2\Gamma((n-1)/2)\pi\Psi_{2}^{\ast}(-1,\frac{n}{2},1)},$$
where $\phi(\cdot)$ is pdf of $1$-dimensional standard normal distribution.

 When $H(\cdot)=Lo(\cdot)$(the cdf of $1$-dimensional standard logistic) in Example $4.3$, it is an $n$-dimensional skew-logistic distribution. Thus,
\begin{align*}
&\mathbf{\Omega}_{\boldsymbol{q},ij}=\frac{2}{\overline{F}_{\mathbf{Z}}(\boldsymbol{\lambda_{q}})}\bigg\{\frac{c_{n}}{c_{n-2,ij}^{\ast\ast}}\overline{E}^{\boldsymbol{\lambda}_{-ij,\boldsymbol{q}}}_{\mathbf{U}_{-ij}}[Lo(\boldsymbol{\gamma}^{T}\boldsymbol{\tau}_{ij,\boldsymbol{q}})]+\frac{c_{n}}{c_{n-1,i}^{\ast\ast}}\gamma_{j}\overline{E}^{\boldsymbol{\lambda}_{-i,\boldsymbol{q}}}_{\mathbf{V}_{-i}}[lo(\boldsymbol{\gamma}^{T}\boldsymbol{\eta}_{i,\boldsymbol{q}})]\\
&+\frac{c_{n}}{c_{n-1,j}^{\ast\ast}}\gamma_{i}\overline{E}^{\boldsymbol{\lambda}_{-j,\boldsymbol{q}}}_{\mathbf{V}_{-j}}[lo(\boldsymbol{\gamma}^{T}\boldsymbol{\eta}_{j,\boldsymbol{q}})]+\frac{\Psi_{1}^{\ast}(-1,\frac{n}{2}+1,1)}{\Psi_{2}^{\ast}(-1,\frac{n}{2},1)}\gamma_{i}\gamma_{j}\\
&\cdot\overline{E}^{\boldsymbol{\lambda}_{\boldsymbol{q}}}_{\mathbf{X}^{\ast\ast}}\bigg[2\sqrt{2\pi}\Psi_{2}^{\ast}(-1,\frac{1}{2},1)lo^{2}(\boldsymbol{\gamma}^{T}\boldsymbol{X}^{\ast\ast})\left(1+\sqrt{2\pi}\phi(\boldsymbol{\gamma}^{T}\boldsymbol{X}^{\ast\ast})\right)\boldsymbol{\gamma}^{T}\boldsymbol{X}^{\ast\ast}\\
&-lo(\boldsymbol{\gamma}^{T}\boldsymbol{X}^{\ast\ast})\boldsymbol{\gamma}^{T}\boldsymbol{X}^{\ast\ast}\bigg]\bigg\}-MTCE_{\mathbf{Z}}(\boldsymbol{\lambda_{q}})_{i}MTCE_{\mathbf{Z}}(\boldsymbol{\lambda_{q}})_{j},~i\neq j,
\end{align*}
and
\begin{align*}
&\mathbf{\Omega}_{\boldsymbol{q},ii}
=\frac{2}{\overline{F}_{\mathbf{Z}}(\boldsymbol{\lambda_{q}})}\bigg\{\frac{c_{n}}{c_{n-1,i}^{\ast}}\lambda_{i,\boldsymbol{q}}\overline{E}^{\boldsymbol{\lambda}_{-i,\boldsymbol{q}}}_{\boldsymbol{W}_{-i}}[Lo(\boldsymbol{\gamma}^{T}\boldsymbol{\xi}_{i,\boldsymbol{q}})] +\frac{c_{n}}{c_{n-1,i}^{\ast\ast}}\gamma_{i}\overline{E}^{\boldsymbol{\lambda}_{-i,\boldsymbol{q}}}_{\mathbf{V}_{-i}}[lo(\boldsymbol{\gamma}^{T}\boldsymbol{\eta}_{i,\boldsymbol{q}})] \\ &+\frac{\Psi_{1}^{\ast}(-1,\frac{n}{2}+1,1)}{\Psi_{2}^{\ast}(-1,\frac{n}{2},1)}\gamma_{i}^{2}\overline{E}^{\boldsymbol{\lambda}_{\boldsymbol{q}}}_{\mathbf{X}^{\ast\ast}}\bigg[\bigg(2\sqrt{2\pi}\Psi_{2}^{\ast}(-1,\frac{1}{2},1)lo(\boldsymbol{\gamma}^{T}\boldsymbol{X}^{\ast\ast})\left[1+\sqrt{2\pi}\phi(\boldsymbol{\gamma}^{T}\boldsymbol{X}^{\ast\ast})\right]
-1\bigg)\\
&\cdot lo(\boldsymbol{\gamma}^{T}\boldsymbol{X}^{\ast\ast})\boldsymbol{\gamma}^{T}\boldsymbol{X}^{\ast\ast}\bigg]\bigg\}
+\frac{\Psi_{1}^{\ast}(-1,\frac{n}{2},1)}{\Psi_{2}^{\ast}(-1,\frac{n}{2},1)}\frac{\overline{F}_{\mathbf{Z^{\ast}}}(\boldsymbol{\lambda_{q}})}{\overline{F}_{\mathbf{Z}}(\boldsymbol{\lambda_{q}})}-[MTCE_{\mathbf{Z}}(\boldsymbol{\lambda_{q}})_{i}]^{2},
\end{align*}
with
\begin{align*} MTCE_{\mathbf{Z}}(\boldsymbol{\lambda_{q}})_{k}=&\frac{2}{\overline{F}_{\mathbf{Z}}(\boldsymbol{\lambda}_{\boldsymbol{q}})}\bigg\{\frac{c_{n}}{c_{n-1,k}^{\ast}}\overline{E}^{\boldsymbol{\lambda}_{-k,\boldsymbol{q}}}_{\boldsymbol{W}_{-k}}[Lo(\boldsymbol{\gamma}^{T}\boldsymbol{\xi}_{k,\boldsymbol{q}})]\\
&+\gamma_{k}\overline{E}^{\boldsymbol{\lambda_{q}}}_{\boldsymbol{M}}\left[\left(1+\exp\left(-\frac{\boldsymbol{M}^{ T}\boldsymbol{M}}{2}\right)\right)\boldsymbol{lo}(\boldsymbol{\gamma^{T}\boldsymbol{M}})\right]\bigg\},
\end{align*}
 $~k=1,~2,\cdots,n,$
$\mathbf{Z}\sim SLo_{n}(\boldsymbol{0},~\mathbf{I_{n}},~\boldsymbol{\gamma}),$ $\mathbf{Z}^{\ast}\sim GSE_{n}\left(\boldsymbol{0},~\boldsymbol{I_{n}},~\boldsymbol{\gamma},~\overline{G}_{n},~Lo(\cdot)\right)$
and $\boldsymbol{lo}(\cdot)$ is pdf of $1$-dimensional standard logistic distribution.
Furthermore, $\mathbf{X}^{\ast\ast}$, $\mathbf{M}$, $\mathbf{W}_{-k}$, $\mathbf{V}_{-k}$ and $\mathbf{U}_{-ij}$ are the same as those in Example $4.3$.

 When $H(\cdot)=\frac{1}{2}$ in Example $4.3$, we obtain MTCov for  logistic distribution in Landsman et al. (2018):
\begin{align*}
\mathbf{\Omega}_{\boldsymbol{q},ij}=&\frac{1}{\overline{F}_{\mathbf{Z}}(\boldsymbol{\lambda_{q}})}\bigg\{\frac{c_{n}}{c_{n-2,ij}^{\ast\ast}}\overline{F}_{\mathbf{U}_{-ij}}(\boldsymbol{\lambda}_{-ij,\boldsymbol{q}})\bigg\}-MTCE_{\mathbf{Z}}(\boldsymbol{\lambda_{q}})_{i}MTCE_{\mathbf{Z}}(\boldsymbol{\lambda_{q}})_{j},~i\neq j,
\end{align*}
and
\begin{align*}
&\mathbf{\Omega}_{\boldsymbol{q},ii}\\
&=\frac{1}{\overline{F}_{\mathbf{Z}}(\boldsymbol{\lambda_{q}})}\bigg\{\frac{c_{n}}{c_{n-1,i}^{\ast}}\lambda_{i,\boldsymbol{q}}\overline{F}_{\boldsymbol{W}_{-i}}(\boldsymbol{\lambda}_{-i,\boldsymbol{q}}) \bigg\}
+\frac{\Psi_{1}^{\ast}(-1,\frac{n}{2},1)}{\Psi_{2}^{\ast}(-1,\frac{n}{2},1)}\frac{\overline{F}_{\mathbf{Z^{\ast}}}(\boldsymbol{\lambda_{q}})}{\overline{F}_{\mathbf{Z}}(\boldsymbol{\lambda_{q}})}-[MTCE_{\mathbf{Z}}(\boldsymbol{\lambda_{q}})_{i}]^{2},
\end{align*}
with
\begin{align*} &MTCE_{\mathbf{Z}}(\boldsymbol{\lambda_{q}})_{k}=\frac{1}{\overline{F}_{\mathbf{Z}}(\boldsymbol{\lambda}_{\boldsymbol{q}})}\bigg\{\frac{c_{n}}{c_{n-1,k}^{\ast}}\overline{F}_{\boldsymbol{W}_{-k}}(\boldsymbol{\lambda}_{-k,\boldsymbol{q}})\bigg\},
\end{align*}
$i,~j,~k=1,~2,\cdots,n$, $\mathbf{Z}\sim Lo_{n}(\boldsymbol{0},~\mathbf{I_{n}})$ and $\mathbf{Z}^{\ast}\sim E_{n}\left(\boldsymbol{0},~\boldsymbol{I_{n}},~\overline{G}_{n}\right)$. In addition, $\mathbf{U}_{-ij}$ and $\mathbf{W}_{-k}$ are the same as those in Example $4.3$.\\
$\mathbf{Example~4.4}$ (Generalized skew-Laplace distribution). The density function of a
generalized skew-Laplace random vector $\mathbf{Y}$, with location parameter $\boldsymbol{\mu}$, scale
matrix $\mathbf{\Sigma}$ and skewing function $H(\cdot): \mathbb{R}\rightarrow\mathbb{R}$, is
given by
\begin{align*}
f_{\boldsymbol{Y}}(\boldsymbol{y})=&\frac{2c_{n}}{\sqrt{|\mathbf{\Sigma}|}}\exp\left\{-[(\boldsymbol{y}-\boldsymbol{\mu})^{T}\mathbf{\Sigma}^{-1}(\boldsymbol{y}-\boldsymbol{\mu})]^{1/2}\right\}H\left(\boldsymbol{\gamma}^{T}\mathbf{\Sigma}^{-\frac{1}{2}}(\boldsymbol{y}-\boldsymbol{\mu})\right),~\boldsymbol{y}\in\mathbb{R}^{n},
 \end{align*}
 where $\boldsymbol{\gamma}=(\gamma_{1},~\gamma_{2},~\cdots,~\gamma_{n})^{T}$ and $c_{n}=\frac{\Gamma(n/2)}{2\pi^{n/2}\Gamma(n)}$.
    We denote it by\\ $\mathbf{Y}\sim GSLa_{n}(\boldsymbol{\mu},~\mathbf{\Sigma},~\boldsymbol{\gamma},~H(\cdot))$. In this case, \\ $g_{n}(u)=\exp\{-\sqrt{2u}\}$, and so
 $$\overline{G}_{n}(t)=(1+\sqrt{2t})\exp(-\sqrt{2t}),$$
 $$\overline{\mathcal{G}}_{n}(t)=(3+2t+3\sqrt{2t})\exp(-\sqrt{2t}).$$
 In addition,
$$c_{n}^{\ast}=\frac{n\Gamma(n/2)}{2\pi^{n/2}\Gamma(n+2)},\;\;
 c_{n}^{\ast\ast}=\frac{n(n+2)\Gamma(n/2)}{2\pi^{n/2}\Gamma(n+4)}.$$ And $H\left(\mathbf{\Sigma}^{-\frac{1}{2}}(\boldsymbol{y}-\boldsymbol{\mu})\right)=H\left(\boldsymbol{\gamma}^{T}\mathbf{\Sigma}^{-\frac{1}{2}}(\boldsymbol{y}-\boldsymbol{\mu})\right)$.\\
 Since
 \begin{align*}
f_{\mathbf{W}_{-k}}(\boldsymbol{t})=c_{n-1,k}^{\ast}\left(1+\sqrt{\boldsymbol{t}^{T}\boldsymbol{t}+\lambda_{k,\boldsymbol{q}}^{2}}\right)\exp\left\{-\sqrt{\boldsymbol{t}^{T}\boldsymbol{t}+\lambda_{k,\boldsymbol{q}}^{2}}\right\},~\boldsymbol{t}\in \mathbb{R}^{n-1},
\end{align*}
 $$f_{\mathbf{V}_{-k}}(\boldsymbol{v})=c_{n-1,k}^{\ast\ast}\left[\left(1+\frac{3}{\sqrt{2}}\right)\left(\boldsymbol{v}^{T}\boldsymbol{v}+\lambda_{k,\boldsymbol{q}}^{2}\right)+3\right]\exp(-\sqrt{\boldsymbol{v}^{T}\boldsymbol{v}+\lambda_{k,\boldsymbol{q}}^{2}}),~\boldsymbol{v}\in \mathbb{R}^{n-1}$$
 and
\begin{align*}
f_{\mathbf{U}_{-ij}}(\boldsymbol{u})=&c_{n-2,ij}^{\ast\ast}\left[\left(1+\frac{3}{\sqrt{2}}\right)\left(\boldsymbol{u}^{T}\boldsymbol{u}+\lambda_{i,\boldsymbol{q}}^{2}+\lambda_{j,\boldsymbol{q}}^{2}\right)+3\right]\\
&\cdot\exp(-\sqrt{\boldsymbol{u}^{T}\boldsymbol{u}+\lambda_{i,\boldsymbol{q}}^{2}+\lambda_{j,\boldsymbol{q}}^{2}}),~\boldsymbol{u}\in \mathbb{R}^{n-2},
\end{align*}
thus
$$c_{n-1,k}^{\ast}=\frac{\Gamma\left(\frac{n-1}{2}\right)}{(2\pi)^{(n-1)/2}}\left[\int_{0}^{\infty}t^{\frac{n-3}{2}}\left(1+\sqrt{2t+\lambda_{k,\boldsymbol{q}}^{2}}\right)\exp\left\{-\sqrt{2t+\lambda_{k,\boldsymbol{q}}^{2}}\right\}\mathrm{d}t\right]^{-1},$$
\begin{align*}
c_{n-1,k}^{\ast\ast}=&\frac{\Gamma\left(\frac{n-1}{2}\right)}{(2\pi)^{(n-1)/2}}\bigg\{\int_{0}^{\infty}t^{\frac{n-3}{2}}\left[(2+3\sqrt{2})\left(t+\frac{\lambda_{k,\boldsymbol{q}}^{2}}{2}\right)+3\right]\\
&\cdot\exp\left(-\sqrt{2t+\lambda_{k,\boldsymbol{q}}^{2}}\right)\mathrm{d}t\bigg\}^{-1}
\end{align*}
and
\begin{align*}
c_{n-2,ij}^{\ast\ast}=&\frac{\Gamma\left(\frac{n-2}{2}\right)}{(2\pi)^{(n-2)/2}}\bigg\{\int_{0}^{\infty}t^{\frac{n-4}{2}}\left[(2+3\sqrt{2})\left(t+\frac{\lambda_{i,\boldsymbol{q}}^{2}}{2}+\frac{\lambda_{j,\boldsymbol{q}}^{2}}{2}\right)+3\right]\\
&\cdot\exp\left(-\sqrt{2t+\lambda_{i,\boldsymbol{q}}^{2}+\lambda_{j,\boldsymbol{q}}^{2}}\right)\mathrm{d}t\bigg\}^{-1}.
\end{align*}
Then
\begin{align*}
\mathbf{\Omega}_{\boldsymbol{q},ij}=&\frac{2}{\overline{F}_{\mathbf{Z}}(\boldsymbol{\lambda_{q}})}\bigg\{\frac{c_{n}}{c_{n-2,ij}^{\ast\ast}}\overline{E}^{\boldsymbol{\lambda}_{-ij,\boldsymbol{q}}}_{\mathbf{U}_{-ij}}[H(\boldsymbol{\gamma}^{T}\boldsymbol{\tau}_{ij,\boldsymbol{q}})]+\frac{c_{n}}{c_{n-1,i}^{\ast\ast}}\gamma_{j}\overline{E}^{\boldsymbol{\lambda}_{-i,\boldsymbol{q}}}_{\mathbf{V}_{-i}}[H'(\boldsymbol{\gamma}^{T}\boldsymbol{\eta}_{i,\boldsymbol{q}})]\\
&+\frac{c_{n}}{c_{n-1,j}^{\ast\ast}}\gamma_{i}\overline{E}^{\boldsymbol{\lambda}_{-j,\boldsymbol{q}}}_{\mathbf{V}_{-j}}[H'(\boldsymbol{\gamma}^{T}\boldsymbol{\eta}_{j,\boldsymbol{q}})]+(n+3)(n+1)\gamma_{i}\gamma_{j}\overline{E}^{\boldsymbol{\lambda}_{\boldsymbol{q}}}_{\mathbf{X}^{\ast\ast}}[H''(\boldsymbol{\gamma}^{T}\boldsymbol{X}^{\ast\ast})]\bigg\}\\
&-MTCE_{\mathbf{Z}}(\boldsymbol{\lambda_{q}})_{i}MTCE_{\mathbf{Z}}(\boldsymbol{\lambda_{q}})_{j},~i\neq j,
\end{align*}
and
\begin{align*}
\mathbf{\Omega}_{\boldsymbol{q},ii}
&=\frac{2}{\overline{F}_{\mathbf{Z}}(\boldsymbol{\lambda_{q}})}\bigg\{\frac{c_{n}}{c_{n-1,i}^{\ast}}\lambda_{i,\boldsymbol{q}}\overline{E}^{\boldsymbol{\lambda}_{-i,\boldsymbol{q}}}_{\boldsymbol{W}_{-i}}[H(\boldsymbol{\gamma}^{T}\boldsymbol{\xi}_{i,\boldsymbol{q}})] +\frac{c_{n}}{c_{n-1,i}^{\ast\ast}}\gamma_{i}\overline{E}^{\boldsymbol{\lambda}_{-i,\boldsymbol{q}}}_{\mathbf{V}_{-i}}[H'(\boldsymbol{\gamma}^{T}\boldsymbol{\eta}_{i,\boldsymbol{q}})] \\ &+(n+3)(n+1)\gamma_{i}^{2}\overline{E}^{\boldsymbol{\lambda}_{\boldsymbol{q}}}_{\mathbf{X}^{\ast\ast}}[H''(\boldsymbol{\gamma}^{T}\boldsymbol{X}^{\ast\ast})]\bigg\}
+(n+1)\frac{\overline{F}_{\mathbf{Z^{\ast}}}(\boldsymbol{\lambda_{q}})}{\overline{F}_{\mathbf{Z}}(\boldsymbol{\lambda_{q}})}\\
&-[MTCE_{\mathbf{Z}}(\boldsymbol{\lambda_{q}})_{i}]^{2},
\end{align*}
with
\begin{align*} MTCE_{\mathbf{Z}}(\boldsymbol{\lambda_{q}})_{k}=&\frac{2}{\overline{F}_{\mathbf{Z}}(\boldsymbol{\lambda}_{\boldsymbol{q}})}\bigg\{\frac{c_{n}}{c_{n-1,k}^{\ast}}\overline{E}^{\boldsymbol{\lambda}_{-k,\boldsymbol{q}}}_{\boldsymbol{W}_{-k}}[H(\boldsymbol{\gamma}^{T}\boldsymbol{\xi}_{k,\boldsymbol{q}})]\\
&+\gamma_{k}\overline{E}^{\boldsymbol{\lambda_{q}}}_{\boldsymbol{M}}\left[(1+\sqrt{\boldsymbol{M}^{ T}\boldsymbol{M}})H'(\boldsymbol{\gamma}^{T}\boldsymbol{M})\right]\bigg\},
\end{align*}
$i,~j,~k=1,~2,\cdots,n$, $\mathbf{Z}\sim GSLa_{n}(\boldsymbol{0},~\mathbf{I_{n}},~\boldsymbol{\gamma},~H(\cdot)),$ \\$\mathbf{Z}^{\ast}\sim GSE_{n}\left(\boldsymbol{0},~\boldsymbol{I_{n}},~\boldsymbol{\gamma},~\overline{G}_{n},~H(\cdot)\right)$,
$\mathbf{X}^{\ast\ast}\sim E_{n}(\boldsymbol{0},~\mathbf{I_{n}},~\overline{\mathcal{G}}_{n})$ and $\mathbf{M}\sim La_{n}(\boldsymbol{0},~\mathbf{I_{n}})$ (Laplace distribution).\\
As for $\frac{c_{n}}{c_{n-1,k}^{\ast}}$, $\frac{c_{n}}{c_{n-1,k}^{\ast\ast}}$ and $\frac{c_{n}}{c_{n-2,ij}^{\ast\ast}}$, we can simplify them further as follows:
$$\frac{c_{n}}{c_{n-1,k}^{\ast}}=\frac{\Gamma\left(\frac{n}{2}\right)2^{(n-3)/2}}{\Gamma(n)\Gamma\left(\frac{n-1}{2}\right)\sqrt{\pi}}\left[\int_{0}^{\infty}t^{\frac{n-3}{2}}\left(1+\sqrt{2t+\lambda_{k,\boldsymbol{q}}^{2}}\right)\exp\left\{-\sqrt{2t+\lambda_{k,\boldsymbol{q}}^{2}}\right\}\mathrm{d}t\right],$$
\begin{align*}
\frac{c_{n}}{c_{n-1,k}^{\ast\ast}}=&\frac{\Gamma\left(\frac{n}{2}\right)2^{(n-3)/2}}{\Gamma(n)\Gamma\left(\frac{n-1}{2}\right)\sqrt{\pi}}\bigg\{\int_{0}^{\infty}t^{\frac{n-3}{2}}\left[(2+3\sqrt{2})\left(t+\frac{\lambda_{k,\boldsymbol{q}}^{2}}{2}\right)+3\right]\\
&\cdot\exp\left(-\sqrt{2t+\lambda_{k,\boldsymbol{q}}^{2}}\right)\mathrm{d}t\bigg\}
\end{align*}
and
\begin{align*}
\frac{c_{n}}{c_{n-2,ij}^{\ast\ast}}=&\frac{2^{(n-6)/2}}{\Gamma(n-1)\pi}\bigg\{\int_{0}^{\infty}t^{\frac{n-4}{2}}\left[(2+3\sqrt{2})\left(t+\frac{\lambda_{i,\boldsymbol{q}}^{2}}{2}+\frac{\lambda_{j,\boldsymbol{q}}^{2}}{2}\right)+3\right]\\
&\cdot\exp\left(-\sqrt{2t+\lambda_{i,\boldsymbol{q}}^{2}+\lambda_{j,\boldsymbol{q}}^{2}}\right)\mathrm{d}t\bigg\}.
\end{align*}

 When $H(\cdot)=La(\cdot)$(the cdf of $1$-dimensional standard Laplace) in Example $4.4$, it is an $n$-variate skew-Laplace distribution. Thus,
\begin{align*}
&\mathbf{\Omega}_{\boldsymbol{q},ij}=\frac{2}{\overline{F}_{\mathbf{Z}}(\boldsymbol{\lambda_{q}})}\bigg\{\frac{c_{n}}{c_{n-2,ij}^{\ast\ast}}\overline{E}^{\boldsymbol{\lambda}_{-ij,\boldsymbol{q}}}_{\mathbf{U}_{-ij}}[La(\boldsymbol{\gamma}^{T}\boldsymbol{\tau}_{ij,\boldsymbol{q}})]+\frac{c_{n}}{c_{n-1,i}^{\ast\ast}}\gamma_{j}\overline{E}^{\boldsymbol{\lambda}_{-i,\boldsymbol{q}}}_{\mathbf{V}_{-i}}[la(\boldsymbol{\gamma}^{T}\boldsymbol{\eta}_{i,\boldsymbol{q}})]\\
&+\frac{c_{n}}{c_{n-1,j}^{\ast\ast}}\gamma_{i}\overline{E}^{\boldsymbol{\lambda}_{-j,\boldsymbol{q}}}_{\mathbf{V}_{-j}}[la(\boldsymbol{\gamma}^{T}\boldsymbol{\eta}_{j,\boldsymbol{q}})]-(n+3)(n+1)\gamma_{i}\gamma_{j}\overline{E}^{\boldsymbol{\lambda}_{\boldsymbol{q}}}_{\mathbf{X}^{\ast\ast}}\left[\frac{\boldsymbol{\gamma}^{T}\boldsymbol{X}^{\ast\ast}}{|\boldsymbol{\gamma}^{T}\boldsymbol{X}^{\ast\ast}|}la(\boldsymbol{\gamma}^{T}\boldsymbol{X}^{\ast\ast})\right]\bigg\}\\
&-MTCE_{\mathbf{Z}}(\boldsymbol{\lambda_{q}})_{i}MTCE_{\mathbf{Z}}(\boldsymbol{\lambda_{q}})_{j},~i\neq j,
\end{align*}
and
\begin{align*}
\mathbf{\Omega}_{\boldsymbol{q},ii}
&=\frac{2}{\overline{F}_{\mathbf{Z}}(\boldsymbol{\lambda_{q}})}\bigg\{\frac{c_{n}}{c_{n-1,i}^{\ast}}\lambda_{i,\boldsymbol{q}}\overline{E}^{\boldsymbol{\lambda}_{-i,\boldsymbol{q}}}_{\boldsymbol{W}_{-i}}[La(\boldsymbol{\gamma}^{T}\boldsymbol{\xi}_{i,\boldsymbol{q}})] +\frac{c_{n}}{c_{n-1,i}^{\ast\ast}}\gamma_{i}\overline{E}^{\boldsymbol{\lambda}_{-i,\boldsymbol{q}}}_{\mathbf{V}_{-i}}[la(\boldsymbol{\gamma}^{T}\boldsymbol{\eta}_{i,\boldsymbol{q}})] \\ &+(n+3)(n+1)\gamma_{i}^{2}\overline{E}^{\boldsymbol{\lambda}_{\boldsymbol{q}}}_{\mathbf{X}^{\ast\ast}}\left[\frac{\boldsymbol{\gamma}^{T}\boldsymbol{X}^{\ast\ast}}{|\boldsymbol{\gamma}^{T}\boldsymbol{X}^{\ast\ast}|}la(\boldsymbol{\gamma}^{T}\boldsymbol{X}^{\ast\ast})\right]\bigg\}
+(n+1)\frac{\overline{F}_{\mathbf{Z^{\ast}}}(\boldsymbol{\lambda_{q}})}{\overline{F}_{\mathbf{Z}}(\boldsymbol{\lambda_{q}})}\\
&-[MTCE_{\mathbf{Z}}(\boldsymbol{\lambda_{q}})_{i}]^{2},
\end{align*}
with
\begin{align*} MTCE_{\mathbf{Z}}(\boldsymbol{\lambda_{q}})_{k}=&\frac{2}{\overline{F}_{\mathbf{Z}}(\boldsymbol{\lambda}_{\boldsymbol{q}})}\bigg\{\frac{c_{n}}{c_{n-1,k}^{\ast}}\overline{E}^{\boldsymbol{\lambda}_{-k,\boldsymbol{q}}}_{\boldsymbol{W}_{-k}}[La(\boldsymbol{\gamma}^{T}\boldsymbol{\xi}_{k,\boldsymbol{q}})]\\
&+\gamma_{k}\overline{E}^{\boldsymbol{\lambda_{q}}}_{\boldsymbol{M}}\left[(1+\sqrt{\boldsymbol{M}^{ T}\boldsymbol{M}})\boldsymbol{la}(\boldsymbol{\gamma}^{T}\boldsymbol{M})\right]\bigg\},
\end{align*}
$i,~j,~k=1,~2,\cdots,n$,
$\mathbf{Z}\sim SLa_{n}(\boldsymbol{0},~\mathbf{I_{n}},~\boldsymbol{\gamma}),$ $\mathbf{Z}^{\ast}\sim GSE_{n}\left(\boldsymbol{0},~\boldsymbol{I_{n}},~\boldsymbol{\gamma},~\overline{G}_{n},~La(\cdot)\right)$, and $\boldsymbol{la}(\cdot)$ is pdf of $1$-dimensional standard Laplace distribution. $|a|$ denotes absolute value of $a$. Moreover, $\mathbf{X}^{\ast\ast}$, $\mathbf{M}$, $\mathbf{W}_{-k}$, $\mathbf{V}_{-k}$ and $\mathbf{U}_{-ij}$ are the same as those in Example $4.4$.

 When $H(\cdot)=\frac{1}{2}$ in Example $4.4$, we obtain a formula of MTCov for  Laplace distribution in Landsman et al. (2018):
\begin{align*}
\mathbf{\Omega}_{\boldsymbol{q},ij}=&\frac{2}{\overline{F}_{\mathbf{Z}}(\boldsymbol{\lambda_{q}})}\bigg\{\frac{c_{n}}{c_{n-2,ij}^{\ast\ast}}\overline{F}_{\mathbf{U}_{-ij}}(\boldsymbol{\lambda}_{-ij,\boldsymbol{q}})\bigg\}-MTCE_{\mathbf{Z}}(\boldsymbol{\lambda_{q}})_{i}MTCE_{\mathbf{Z}}(\boldsymbol{\lambda_{q}})_{j},~i\neq j,
\end{align*}
and
\begin{align*}
\mathbf{\Omega}_{\boldsymbol{q},ii}
&=\frac{2}{\overline{F}_{\mathbf{Z}}(\boldsymbol{\lambda_{q}})}\bigg\{\frac{c_{n}}{c_{n-1,i}^{\ast}}\lambda_{i,\boldsymbol{q}}\overline{F}_{\boldsymbol{W}_{-i}}(\boldsymbol{\lambda}_{-i,\boldsymbol{q}}) \bigg\}
+(n+1)\frac{\overline{F}_{\mathbf{Z^{\ast}}}(\boldsymbol{\lambda_{q}})}{\overline{F}_{\mathbf{Z}}(\boldsymbol{\lambda_{q}})}-[MTCE_{\mathbf{Z}}(\boldsymbol{\lambda_{q}})_{i}]^{2},
\end{align*}
with
\begin{align*} &MTCE_{\mathbf{Z}}(\boldsymbol{\lambda_{q}})_{k}=\frac{2}{\overline{F}_{\mathbf{Z}}(\boldsymbol{\lambda}_{\boldsymbol{q}})}\bigg\{\frac{c_{n}}{c_{n-1,k}^{\ast}}\overline{F}_{\boldsymbol{W}_{-k}}(\boldsymbol{\lambda}_{-k,\boldsymbol{q}})\bigg\},
\end{align*}
 $i,~j,~k=1,~2,\cdots,n$, $\mathbf{Z}\sim La_{n}(\boldsymbol{0},~\mathbf{I_{n}})$ and $\mathbf{Z}^{\ast}\sim E_{n}\left(\boldsymbol{0},~\boldsymbol{I_{n}},~\overline{G}_{n}\right)$. In addition, $\mathbf{U}_{-ij}$ and $\mathbf{W}_{-k}$ are the same as those in Example $4.4$.
 \section{Numerical illustration}
 We consider MTCov risk measure for the normal (N) and skew-normal (SN) distributions.\\
  Let $\mathbf{P}=(P_{1},P_{2},P_{3})^{T}\sim N_{3}(\boldsymbol{\mu},\mathbf{\Sigma})$ and $\mathbf{Q}=(Q_{1},Q_{2},Q_{3})^{T}\sim SN_{3}(\boldsymbol{\mu},\mathbf{\Sigma},\boldsymbol{\gamma})$, with
\begin{align*}
\boldsymbol{\mu}=\left(\begin{array}{ccccccccccc}
1.3\\
0.8\\
3.2
\end{array}
\right),
\mathbf{\Sigma}=\left(\begin{array}{ccccccccccc}
1.33&-0.067&2.63\\
-0.067&0.25&-0.50\\
2.63&-0.50&5.76
\end{array}
\right)
~and~
\boldsymbol{\gamma}=\left(\begin{array}{ccccccccccc}
2.1\\
-0.045\\
-1.06
\end{array}
\right).
\end{align*}
We let $\boldsymbol{q}=(0.80,0.90,0.95)^{T}$, then the results are presented in Tables 1 and 2, respectively:
\\
\begin{table}[!htbp]
\centering
\begin{tabular}{|c||c|c|c|c|c|c|}
  \hline
  \backslashbox {$i$}{$MTCov_{\mathbf{P},ij}$}{$j$} &$1$& $2$&$3$\\
  \hline
  $1$ &0.2908 & -0.0146& 0.5751\\
  \hline
  $2$&-0.0146 & 0.0400&  -0.0875\\
  \hline
  $3$&0.5751 & -0.0875&  1.2243\\
  \hline
\end{tabular}\\
Table 1: The MTCov of normal distribution for $\boldsymbol{q}=(0.80, 0.90, 0.95)^T$.
\end{table}
\begin{table}[!htbp]
\centering
\begin{tabular}{|c||c|c|c|c|c|c|}
  \hline
  \backslashbox {$i$}{$MTCov_{\mathbf{Q},ij}$}{$j$} &$1$& $2$&$3$\\
  \hline
  $1$ &0.2222 & -0.0131& 0.4429\\
  \hline
  $2$&-0.0131 & 0.0389&  -0.0829\\
  \hline
  $3$&0.4429 & -0.0829&  0.9761\\
  \hline
\end{tabular}\\
Table 2: The MTCov of skew normal distribution for $\boldsymbol{q}=(0.80, 0.90, 0.95)^T$.
\end{table}

 From Tables 1 and 2, we can observe that the main diagonal and sub diagonal MTCov for normal distributions are greater than that for corresponding skewed distribution, and others are opposite.

 In addition, using Tables 1 and 2 and Eq. (\ref{(20)}) we can give their MTCorr matries, respectively:
\begin{align*}
MTCorr_{\mathbf{P}}=\left(\begin{array}{ccccccccccc}
1&-0.135371&0.963834\\
-0.135371&1&-0.3953977\\
0.963834&-0.3953977&1
\end{array}
\right)
\end{align*}
and
\begin{align*}
MTCorr_{\mathbf{Q}}=\left(\begin{array}{ccccccccccc}
1&-0.1409044&0.9510131\\
-0.1409044&1&-0.4254344\\
0.9510131&-0.4254344&1
\end{array}
\right).
\end{align*}
 \section{Concluding remarks}
 This paper has presentd MTCov for generalized skew-elliptical distributions, which is an extending of MTCE for generalized skew-elliptical distributions (Zuo and Yin (2020)). It is not only generalizes MTCov for elliptical distributions (Landsman et al. (2018)), but also extends TV for generalized skew-elliptical distributions (Eini and Khaloozadeh (2020)). For examples, generalized skew-normal, generalized skew student-$t$, generalized skew-logistic and generalized skew-Laplace distributions are given. To illustrate our results can be computed in the theorems, the numerical illustrations of the obtained results are given. Note that, in general calculation, we can use Cholesky
decomposition (see Golub and Van Loan (2012), for example), there exists a unique lower triangular matrix $\mathbf{A}$
such that $\mathbf{AA}^{T}=\mathbf{\Sigma}$. Let $\mathbf{Z}=\mathbf{A}^{-1}(\mathbf{Y}-\boldsymbol{\mu})$, similar to Theorem 1, a special formula of the MTCov for generalized skew-elliptical distributions is given by
 $$MTCov_{\mathbf{Y}}(\boldsymbol{y_{q}})=\mathbf{A}\mathbf{\Upsilon}_{\boldsymbol{q}}\mathbf{A}^{T},$$
 where  $\mathbf{\Upsilon}_{\boldsymbol{q}}$ is the same as $\mathbf{\Omega}_{\boldsymbol{q}}$ in Theorem $1$, except that $$\boldsymbol{\lambda_{q}}=\mathbf{A}^{-1}(VaR_{\boldsymbol{q}}(\mathbf{Y})-\boldsymbol{\mu}).$$
 Furthermore, in Kim and Kim (2019), the authors provided expressions of TCE and TV for mean-variance mixture normal distributions. We hope that the result for MTCov can be extended to those mixture distributions in future research.
\section*{Acknowledgments}
\noindent  The research was supported by the National Natural Science Foundation of China (No. 12071251, 11571198, 11701319)
\section*{Conflicts of Interest}
\noindent The authors declare that they have no conflicts of interest.
\section*{References}
\bibliographystyle{model1-num-names}

\begin{thebibliography}{99}
 \bibitem{Adcock2019}Adcock, C., Landsman, Z., Shushi, T., 2019. Stein's lemma for generalized skew-elliptical random vectors. Communications in Statistics-Theory and Methods, 1-16. doi:10.1080/03610926.2019.1678642.

 \bibitem{Denuit2005}Denuit, M., Dhaene, J., Goovaerts, M., Kaas, R., 2005. Actuarial Theory for Dependent Risks : Measures, Orders and Models. John Wiley and Sons, Ltd, West Sussex.

 \bibitem{Eini2020}Eini, E.J., Khaloozadeh, H., 2020. Tail variance for
generalized skew-elliptical distributions. Communications in Statistics-Theory and Methods, 1-18. doi:
10.1080/03610926.2020.1751853.
 \bibitem{Fang1990}Fang, K.T., Kotz, S., Ng, K.W., 1990. Symmetric Multivariate and Related Distributions. Chapman and Hall, New York.

\bibitem{Furman2006}Furman, E., Landsman, Z. M., 2006. Tail variance premium with applications for elliptical
portfolio of risks. ASTIN Bulletin 36 (2), 433-462.

\bibitem{Golub2012}
 Golub, G.H.,   Van Loan, C.F., 2012. Matrix Computation (4th Edition). Johns Hopkins University Press, Baltimore, Maryland.

\bibitem{Kim2019}Kim, J.H.T., Kim, S.Y., 2019. Tail risk measures and risk allocation for the class of multivariate normal mean-variance mixture distributions. Insurance: Mathematics and Economics 86, 145-157.



\bibitem{Landsman2016}Landsman, Z., Makov, U., Shushi, T., 2016. Multivariate tail conditional expectation for elliptical distributions. Insurance: Mathematics and Economics 70, 216-223.
\bibitem{Landsman2018}
Landsman, Z., Makov, U.,  Shushi, T., 2018. A multivariate tail covariance measure for elliptical distributions. Insurance: Mathematics and Economics 81, 27-35.
\bibitem{Lin2006}
Lin, S.D., Srivastava, H. M., Wang, P. Y., 2006. Some expansion formulas for a class of generalized Hurwitz-Lerch Zeta functions. Integral
Transforms and Special Functions 17 (11), 817-827.

\bibitem{McNeil2005}McNeil, A.J., Frey, R., Embrechts, P., 2005. Quantitative Risk Management: Concepts, Techniques
and Tools. Princeton University Press, New Jersey.
\bibitem{Mousavi2019}Mousavi, S.A., Amirzadeh, V., Rezapour, M., Sheikhy, A., 2019. Multivariate tail conditional expectation for scale mixtures of skew-normal distribution. Journal of Statistical Computation and Simulation, 1-15. doi:10.1080/00949655.2019.1657864.
\bibitem{Zuo2020}Zuo, B., Yin, C., 2020. Tail conditional expectations for generalized skew-elliptical distributions. Probability in the Engineering and Informational Sciences,1-14.
doi:10.1017/S0269964820000674.
\bibitem{Zuo2021}Zuo, B., Yin, C., Balakrishnan, N., 2021.
Expressions for joint moments of elliptical distributions.
Journal of Computational and Applied Mathematics
391 (2021), 113418.













\end{thebibliography}







\end{document}